\documentclass[preprint, 12pt]{elsarticle}
\usepackage{lineno,hyperref,amssymb,amsmath,booktabs,subfigure}
\usepackage{comment}
\usepackage{graphicx}
\usepackage{siunitx}
\usepackage[dvipsnames]{xcolor}
\biboptions{numbers,sort&compress}

\journal{International Journal of Heat and Mass Transfer}

\newcommand{\bi}{ \boldsymbol}
\newcommand\Figref[1]{\hyperref[#1]{Figure~\ref*{#1}}}
\newcommand\Tabref[1]{\hyperref[#1]{Tab.~\ref*{#1}}}
\newcommand\Eqref[1]{\hyperref[#1]{Eq.~(\ref*{#1})}}
\newcommand\Sectref[1]{\hyperref[#1]{§\ref*{#1}}}

\begin{document}
\begin{frontmatter}
\title{Layer coupling between solutal and thermal convection in liquid
metal batteries}

\author[hzdr]{Paolo Personnettaz\fnref{fn1}\corref{cor1}}
\author[hzdr,oldenburg]{Tanja Sophia Klopper\fnref{fn1}}
\author[hzdr]{Norbert Weber}
\author[hzdr]{Tom Weier}

\address[hzdr]{Helmholtz-Zentrum Dresden -- Rossendorf, Bautzner Landstr.\ 400, 01328 Dresden, Germany}
\address[oldenburg]{Carl von Ossietzky University Oldenburg, Ammerländer Heerstr. 114-118, 26129 Oldenburg, Germany}

\cortext[cor1]{Corresponding author. Helmholtz-Zentrum Dresden –
  Rossendorf, Bautzner Landstr. 400, 01328 Dresden,
  Germany. \textit{E-mail address}: p.personnettaz@hzdr.de (P. Personnettaz)}
\fntext[fn1]{The two first authors contributed equally to the publication.}

\begin{abstract}
    For longer than one decade, liquid metal batteries (LMBs) are
  developed with the primary aim to provide economic stationary energy
  storage. Featuring two liquid metal electrodes separated by a molten
  salt electrolyte, LMBs operate at elevated temperature as simple
  concentration cells. Therefore, efficient mass transfer is a basic
  prerequisite for their economic operation.
  Understanding these mechanisms cannot be limited at the single layer level.
    With this motivation, the effects of solutal- and
  thermally-driven flow are studied, as well as the flow coupling
  between the three liquid layers of the cell. It is shown that
  solutal convection appears first and thermal convection much later.
    While the presence of solutal flow depends on the mode
  of operation (charge or discharge), the occurrence of thermal
  convection is dictated by the geometry (thickness of layers).
    The coupling of the flow phenomena between the
  layers is intriguing: while thermal convection is
  confined to its area of origin, i.e. the electrolyte, solutal
  convection is able to drive flow in the positive electrode
  and in the electrolyte.
\end{abstract}

\begin{highlights}
\item thermal convection appears first in the electrolyte -- at charge and discharge
\item solutal convection emerges much faster and stronger than thermal convection
\item solutal convection can mix all layers by viscous coupling, thermal convection not
\item solutal convection affects internally heated convection in the electrolyte
\item coupled top interface allows for better mixing in the positive electrode than no-slip
\end{highlights}

\end{frontmatter}

\section{Introduction}
\begin{figure}[hbt]
\centering
\includegraphics[width=0.5\textwidth]{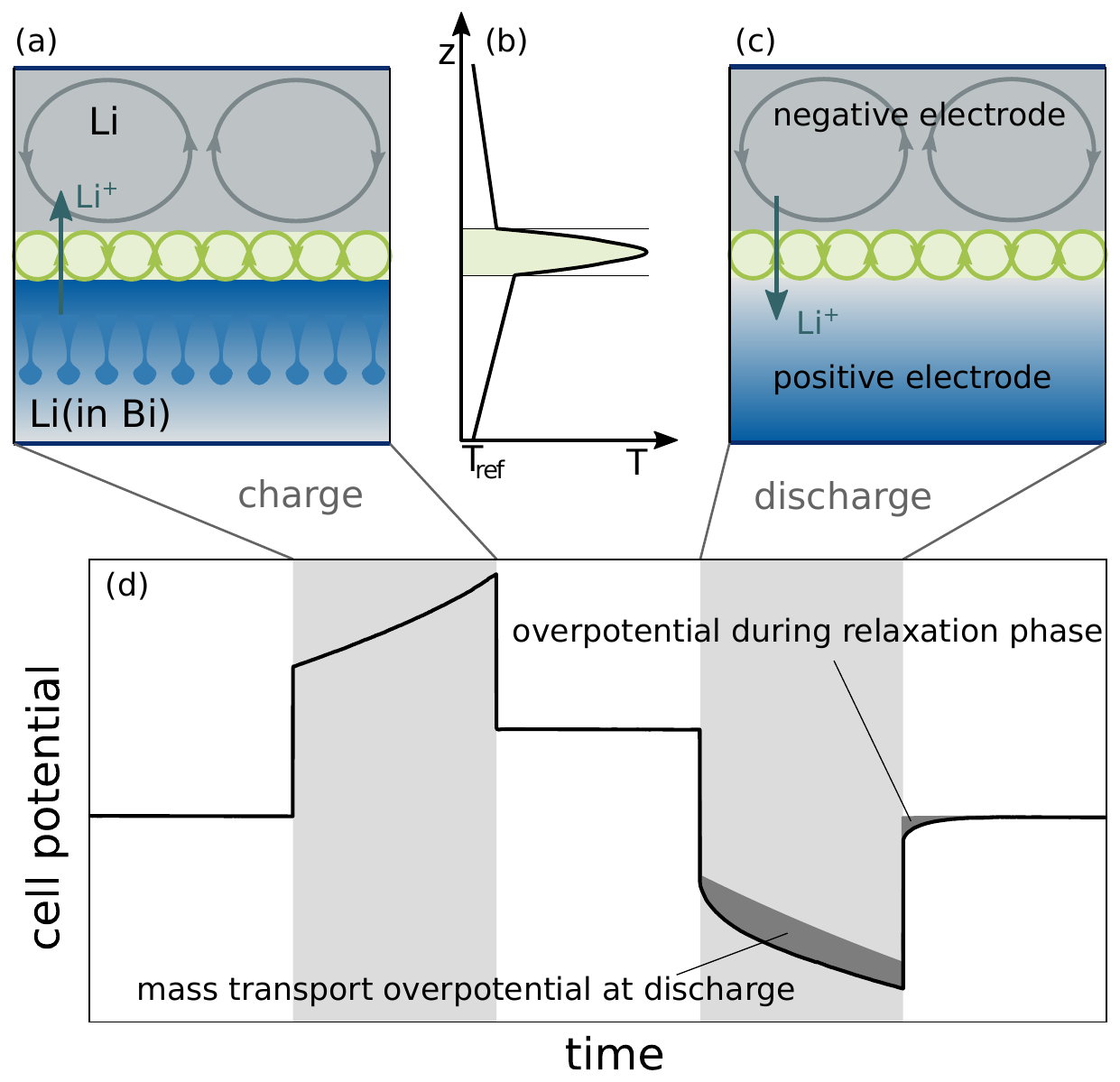}
\caption{Setup and operation of a Li$||$Bi LMB at charge (a) and
  discharge (c) as well as the typical vertical temperature profile
  during operation (b). Sub-figure (d) illustrates a charge-discharge cycle
	with typical effects due to mass-transport overvoltage \cite{Personnettaz2019}.}
\label{f:setup}
\end{figure}

Since the turn of the millennium, liquid metal batteries (LMBs) are
discussed as low-cost stationary electrical energy storage for fluctuating renewable sources.
\Figref{f:setup} illustrates the typical setup of
such a device. The negative electrode on top is separated from
the positive electrode by a thin molten salt electrolyte layer \cite{Kim2013a}. In order to
prevent a short-circuit between the electrodes, violent fluid flow needs to be avoided
\cite{Kelley2018}.

LMBs are simple concentration cells. At discharge, the electro-active species migrates
from the negative electrode through the electrolyte before being alloyed into the positive electrode.
It has already been noticed during the first days of LMB research that non-ideal mixing during this alloying process leads to mass transport overpotentials.
A highly concentrated solute region located near the
interfaces will develop especially at high current density \cite{Agruss1962a}. In extreme cases, even solid
intermetallic phases might form at the electrolyte-positive electrode interface
\cite{Cairns1967,Vogel1967}. Similarly early, thermal convection
\cite{Vogel1967} and mechanical stirring \cite{Cairns1967,Foster1967b}
have been proposed to mitigate concentration overpotentials.

In recent years, many flow effects have been investigated aiming to
enhance mass transfer in liquid metal electrodes.
Electro-vortex flow (EVF), which is driven by converging or diverging current lines, has been explored
by Ashour et al.~\cite{Ashour2017a}. Shortly later, it has been shown
that the placement of the feeding lines as well as stray magnetic
fields have a critical influence on the flow structure and magnitude
of EVF \cite{Weber2018}. Using an energy balance and numerical
modelling, Herreman et al.~concluded that non-swirling electro-vortex
flow is too weak to break the stable density stratification during discharge inside the
positive electrode \cite{Herreman2020a}.

Recently, Weber et al.~combined convection with an electrochemical model showing that the
influence of EVF on the cell voltage is indeed limited. On the other
hand, a vertical magnetic background field, which drives strong
swirling flow, is able to increase the cell efficiency substantially
\cite{Weber2020}.
The latter has been confirmed theoretically and with DNS simulations by Herreman et al.~as well  \cite{Herreman2021}.

Thermal convection in LMBs has been subject of various investigations.
Motivated by mixing the positive electrode alloy, Kelley \& Sadoway
heated a PbBi layer from below measuring the flow velocity
\cite{Kelley2014}; the experiment was later reproduced numerically by
Beltr{\'a}n et al. \cite{Beltran2016}. Shen \& Zikanov were the first who
modelled thermal convection in a complete LMB. They show that it is caused
mainly by the heat released in the electrolyte, as illustrated in
\Figref{f:setup}b. Consequently, flow is expected to appear in
the negative electrode and electrolyte, while the positive electrode is stably stratified.
Flow structures and magnitude depend very much on the thickness of the
single layers \cite{Shen2015}.
Köllner et al.~\cite{Koellner2017} confirmed these early investigations and
pointed out the presence of three additional thermally driven flows:
Marangoni flows at both interfaces and anti-convection in the lower layer.

Finally, Personnettaz et al. discussed the presence of electrochemical heat, and
highlighted that in realistic LMBs with thin electrolyte layers (5\,mm)
thermal convection will be dominant in the thick negative electrode, but negligible in the
electrolyte itself \cite{Personnettaz2018a}.

While thermal convection and EVF appear independently of the direction
of the current, solutal convection can be observed only at charge. As
illustrated in \Figref{f:setup}a for a Li$||$Bi cell, Li is
transferred then from the positive electrode alloy towards the negative electrode.
The heavy (Li-poor) Bi-Li alloy will sink down in plumes.
At discharge, this process is reversed:
Li is alloyed on top of the Bi-rich layer forming a stable density
stratification (\Figref{f:setup}c). This ``asymmetry'' of the
effects leads to an ``asymmetry'' within the corresponding
charge-discharge curve, shown in \Figref{f:setup}d. At discharge a
considerable mass-transport overpotential is observed, which is
not present while charging. When switching off the current, the
voltage returns immediately to the equilibrium potential after a charging
phase (solutal flow), but needs a certain diffusive relaxation time after a
discharge phase (stable density stratification). The crucial relevance
of solutal convection and the stable density stratification
has to our best knowledge first been discussed by Kelley \& Weier
\cite{Kelley2018} and has for the first time been evidenced by
Personnettaz et al.~\cite{Personnettaz2019}. The findings have later
been confirmed by Herreman et al.~\cite{Herreman2020a} performing
two- and three-dimensional numerical simulations of mass transport and EVF.
Interestingly, the two mass-transfer polarization effects, highlighted in \Figref{f:setup}d can also be observed in
older measurements. For example, the charge-discharge curve of a
Na$||$Bi cell from 2015 shows an additional overpotential at
discharge \cite{Stefani2015}, and Fig. 8b in Kim et al.~\cite{Kim2013b}
the relaxation of the cell potential after a discharge
cycle. Tab.~\ref{t:polarization} gives a few more examples of related
effects in the old literature.
\begin{table}[h]
\caption{Experimental results showing mass transport overvoltage in the
	literature of liquid metal electrodes. Chronological order.}\label{t:polarization}
\centering
\footnotesize
\begin{tabular}{ll}
\toprule
	\multicolumn{1}{c}{observation} & \multicolumn{1}{c}{source}\\
\midrule
	polarization effects at high discharge current in Li-Cd and Li-Zn cells   & \cite[p. 202]{Lawroski1962b}\\
	clear concentration polarization, when switching off current after discharge  & \cite[p. 75]{Crouthamel1967}\\
	strong asymmetry of interface concentration between charge and discharge  & \cite[p. 955]{Shimotake1967}\\
	clear concentration polarization effects after switching off the current  & \cite[p. 159]{Bradwell2011}\\
	charge-discharge curve shows additional (concentration) losses at discharge  &  \cite{Stefani2015}\\
\bottomrule
\end{tabular}
\end{table}
The aim of this article is to elucidate the interaction between solutal and
thermal convection in particular during the transient start-up of the
battery.
In this context, we pay special attention to the coupling of
the flow between the three distinct layers.
In the first part of the article, we state the relevant equations that describe
thermal and solutal convection in three-layer LMBs.
From them, we deduce the relevant dimensionless number and we provide
estimations of the expected orders of magnitude for different chemistries.
Previous studies are then used to compute the critical current density at which thermal convection appears in the two top layers.
Then, we move to the numerical modelling of 2D Li$||$Bi cells during charge and discharge.
There, we discuss the effect of sign and magnitude of the current on heat, mass and momentum transfer.

\section{Theory}
\label{s:theory}
The distinct regions $\Omega^{(i)}$ (N-negative electrode, E-electrolyte, P-positive electrode) of an LMB are
coupled at their interfaces $\partial\Omega^{(i|j)}$, see \Figref{f:geometry}.
They permit the exchange of momentum, heat and -- for certain ions -- mass
through electrochemical reactions.
The transport processes inside the regions depend on the boundary conditions and material properties.
These vary widely from layer to layer and so do the transport phenomena.
In this section we will discuss the equations describing these mechanisms, we will rearrange the relevant parameters in the form of dimensionless numbers and
provide indications on the presence of convection in the different layers.
\subsection{Mass transport}
\label{s:massTransport}
Mass transport is vital for the operation of electrochemical cells.
In the negative electrode of LMBs, concentration gradients are typically absent because it consists in many cases of a pure substance (but see Ca-Bi$||$Ca-Sb \cite{Ouchi2014}, Ca-Mg$||$Bi \cite{Ouchi2016} and Li-Mg$||$Sb-Pb \cite{Blanchard2013} for counterexamples).
In this work, we neglect the presence of advective and diffusive transport of ions in the electrolyte.
Assuming a perfectly blended electrolyte seems justified by experimental evidence \cite{Agruss1963}.
Therefore, concentration distributions are considered only in the positive electrode, assuming a binary mixture A-B (in the present study Li-Bi).
The mass concentration of component A, $\rho_\mathrm{A}$ (\si{\kilo\gram\per\cubic\meter}), is described by the advection-diffusion equation
\cite{bird2002transportphenomena,Personnettaz2019,Herreman2020a}
\begin{equation}
	\frac{\partial \rho_\mathrm{A}}{\partial t} + \mathbf{u} \cdot \nabla \rho_\mathrm{A}
     = \mathcal{D}_\mathrm{AB}  \nabla^2 \rho_\mathrm{A} \ \mathrm{in}
     \  \Omega^\mathrm{(P)},
	\label{eq:AdvectionDiffusionMass}
\end{equation}
in which $\mathbf{u}$ is the velocity field and
$\mathcal{D}_\mathrm{AB}$ is the mass transport diffusion coefficient.
We neglect the volume change of the positive electrode due to the density variation and concentration dependent mass transport
coefficient \cite{weber2021cell}.
As the solid boundaries are impermeable, we impose an homogeneous Neumann condition for the concentration.
The interface between the electrolyte and the positive electrode allows for the
transfer of mass, because ions are reduced or oxidised there as $\mathrm{A}^+ + \mathrm{e}^- \leftrightharpoons  \mathrm{A(in B)}$
through electrochemical reactions.
The mass flux of the component A under amperostatic condition, $\dot{m}''_\mathrm{A}$, can be computed with the Faraday law of electrolysis.
The resulting boundary condition can be formulated then as
\begin{equation}
	-\mathcal{D}_\mathrm{AB} \nabla \rho_\mathrm{A}  \cdot \textbf{n} =
    \dot{m}''_\mathrm{A} = \frac{j
    \mathcal{M}_\mathrm{A}}{n_\mathrm{el}\mathrm{F}} \ \mathrm{at} \
    \partial\Omega^\mathrm{(\mathrm{P|E})} \ ,
	\label{eq:FaradyMassFluxBC}
\end{equation}
where $j$, $\mathcal{M}_\mathrm{A}$, $\textbf{n}$, $n_\mathrm{el}$, $\mathrm{F}$ are
current density, molar mass of the component A (Li in our application), the surface normal vector, the number of charges exchanged by the electrochemical reaction and the Faraday constant, respectively.
We consider only the primary current distribution, which is uniform across all
the domain, assuming that the top and bottom plate are ideal current collectors and that
the lateral walls are insulated.
\subsection{Heat transport}
\label{s:heatTransport}
In LMBs, heat is mainly generated by the passage of the current through the
electrolyte, and then transported into the full cell.
For an overview of thermal phenomena in LMB and the relevant assumptions, please refer
to Personnettaz et al.~\cite{Personnettaz2018a}.
The transport equation for the temperature $T$ in the $i$th layer reads
\begin{equation}
	\label{eq:AdvectionDiffusionTemperature}
	\frac{\partial T}{\partial t} + \mathbf{u}\cdot\nabla T = \mathcal{D}^{(i)}_{T} \nabla^2 T + \frac{\mathcal{D}_\mathrm{T}^{(i)}}{k^{(i)}} (\dot{q}''')^{(i)} \  \mathrm{in} \ \Omega^{(i)} \ \mathrm{for} \ i= \mathrm{N, E, P},
\end{equation}
in which $\mathcal{D}_\mathrm{T}$ and $k$ are thermal diffusivity and conductivity.
The ohmic internal heat generation is described by $\dot{q}'''=\rho_\mathrm{el}j^2$, in which $\rho_\mathrm{el}$ is the electrical resistivity. Electrochemical heat generation is completely neglected.
The top and bottom boundary are held at uniform temperature $T_\mathrm{ref}$.
Lateral sides are assumed adiabatic.
At the interfaces between the layers, $\partial\Omega^{(i|j)}$, continuity of temperature and heat flux must be enforced.
Most of these assumptions are shared by previous works \cite{Personnettaz2018a, Koellner2017, Shen2015}.

\subsection{Fluid Flow}
\label{s:fluidFlow}
In LMBs we have three immiscible liquid phases; however, the flow in each layer is characterised by an independent topology.
We work in the assumption of fixed interfaces and of the Boussinesq approximation, similar as other authors \cite{Koellner2017, Shen2015}.
The momentum balance is defined as
\begin{multline}
	\label{eq:MomentumBoussinesq}
	\frac{\partial\mathbf{u}}{\partial t}+\mathbf{u}\cdot\nabla\mathbf{u} = -\frac{1}{\rho_\mathrm{ref}^{(i)}}\nabla p_\mathrm{d} +\nu^{(i)}\nabla^2\mathbf{u} \\
	+ (\beta_\mathrm{T}^{(i)} (T - T_\mathrm{ref}) +  \beta_{\rho_\mathrm{A}}^{(k)} (\rho_\mathrm{A} - \rho_\mathrm{A,ref})) \mathbf{g}  \  \mathrm{in} \ \Omega^{(i)} \ \mathrm{for} \ i= \mathrm{N, E, P} \ \mathrm{and} \ k= \mathrm{P} ,
\end{multline}
where $\rho_{\mathrm{ref}}$ is the density in the reference thermodynamic condition, $\nu$ is the kinematic viscosity, $p_\mathrm{d}$ is the modified pressure,  $\mathbf{g}$ is the gravity acceleration vector and  $\beta_\mathrm{T}$ and $\beta_{\rho_\mathrm{A}}$ are thermal and solutal expansion coefficient \cite{leal2007advanced}.
We neglect the presence of other body forces, like the one generated by magneto-hydrodynamic effects.
The velocity field satisfies the continuity equation in the incompressible form $\nabla\cdot\mathbf{u} = 0$.
At the interfaces, $\partial\Omega^{(i|j)}$, the normal velocity is enforced to be zero.
The tangential components are assumed continuous across the interfaces.
Furthermore, the stress balance
\begin{equation}
	(\rho_\mathrm{ref}\nu)^{\mathrm{(i)}}\frac{\partial u_{k}}{\partial z} = (\rho_\mathrm{ref}\nu)^{\mathrm{(E)}}\frac{\partial u_{k}}{\partial z} \  \mathrm{at} \ \partial\Omega^{(i|\mathrm{E})} \ \mathrm{for} \ i = \mathrm{N, P}  \ \mathrm{and} \ k = x, y
	\label{eq:stressEquilibriumSimpleInterface}
\end{equation}
must be satisfied. In this context, solutal and thermal Marangoni effects are neglected.
The walls are rigid and impermeable and described by a no-slip boundary condition.

\subsection{Dimensionless numbers}
The system of equations that describes convective heat and mass transport in a
3-layer LMB contains 14 different material properties, the reference state ($T_\mathrm{ref}$, $\rho_\mathrm{A,ref}$),
and one operating parameter: the current density $j$.
Furthermore, we have to consider the geometrical dimensions: three layer heights $\Delta h^\mathrm{(i)}$ and the
lateral extension of the cell $L_{x}$.
This large number of parameters motivates performing a dimensionless analysis.
Here, we consider every layer per se and we use as example the positive electrode in
which both heat and mass transfer take place.
At least five distinct time scales can be defined: three based on diffusion ($\mathcal{T}_{\mathcal{D}i}=\mathcal{L}^2\mathcal{D}_i^{-1}$)
and two based on the free fall velocity (e.g. $\mathcal{T}_{\beta \Theta} =
\mathcal{L}(g\beta_\mathrm{T}\Delta\Theta\mathcal{L})^{-0.5} $), in which
$\mathcal{D}_i$, $\mathcal{L}$ and $\Delta \Theta$ are the diffusivity of the
$i$-th component, the reference height and temperature difference, respectively.
If we substitute them in the system of equations, the following five dimensionless numbers appear \cite{bird2002transportphenomena}:
\begin{align}
	\text{Prandtl number:\qquad }Pr                      & =
	\frac{\nu}{\mathcal{D}_\mathrm{T}} & = \frac{\mathcal{T}_{\mathcal{D}\Theta}}{\mathcal{T}_{\mathcal{D}u}},                                                          \\
	\text{Schmidt number:\qquad }Sc                      & =
	\frac{\nu}{\mathcal{D}_\mathrm{AB}} & = \frac{\mathcal{T}_{\mathcal{D}\mathcal{C}}}{\mathcal{T}_{\mathcal{D}u}},                                                         \\
	\text{Lewis number:\qquad }Le                        & =
	\frac{\mathcal{D}_\mathrm{T}}{\mathcal{D}_\mathrm{AB}} & = \frac{\mathcal{T}_{\mathcal{D}\mathcal{C}}}{\mathcal{T}_{\mathcal{D}\Theta}},                                      \\
	\text{Thermal Rayleigh number:\qquad }Ra_{\Theta}    & =
	\frac{g\beta_\mathrm{T}\Delta \Theta
		\mathcal{L}^3}{\mathcal{D}_\mathrm{T}\nu}  & =
	\frac{\mathcal{T}_{\mathcal{D}u}\mathcal{T}_{\mathcal{D}\Theta}}{\mathcal{T}_{\beta
			\Theta}^2},   \label{eq:Ra_T}           \\
	\text{Solutal Rayleigh number:\qquad }Ra_\mathcal{C} & =
	\frac{g\beta_{\rho_\mathrm{A}}\Delta \mathcal{C}
		\mathcal{L}^3}{\mathcal{D}_\mathrm{AB}\nu} & =
	\frac{\mathcal{T}_{\mathcal{D}u}\mathcal{T}_{\mathcal{D}\mathcal{C}}}{\mathcal{T}_{\beta\mathcal{C}}^2},  \label{eq:Ra_C} \\
	\text{Density ratio:\qquad }R_\rho                   & =
	\frac{\beta_{\rho_\mathrm{A}}\Delta \mathcal{C}}{\beta_\mathrm{T}\Delta
		\Theta} &=
	\frac{\mathcal{T}_{\beta\Theta}^2}{\mathcal{T}_{\beta\mathcal{C}}^2}.
\end{align}
Here, $\Delta \mathcal{C}$ denotes the reference concentration difference across the layer.
The first three dimensionless numbers are ratios of couples of diffusivities and related time scales.
The two Rayleigh numbers quantify the strength of the buoyancy with respect to dissipative effects of viscosity and component diffusion.
Lewis number and density ratio are not essential parameters of the problem, but they are helpful to understand the relative importance of heat and mass transport.

The concentration difference across the positive electrode can be estimated a priori using the boundary conditions of Eq.~\ref{eq:FaradyMassFluxBC}:
\begin{equation}
    \Delta \mathcal{C}_\mathrm{a priori} =  \frac{\dot{m}''_\mathrm{A} \Delta h^\mathrm{(P)}}{\mathcal{D}_\mathrm{AB}} = \frac{j \mathcal{M}_\mathrm{A}\Delta h^\mathrm{(P)}}{n_\mathrm{el}\mathrm{F} \mathcal{D}_\mathrm{AB}} \ .
    \label{eq:ConcentrationUnitMassFlux}
\end{equation}
This value largely overestimates the real difference, as the concentration gradient is present only in a thin region near the interface.
A better estimation can be done a posteriori, applying a 1D diffusion model, using the semi-infinite approximation:
\begin{equation}
	\Delta \mathcal{C}_\mathcal{D}(t) =   \frac{j\mathcal{M}_\mathrm{A}}{n_\mathrm{el}\mathrm{F} \mathcal{D}_\mathrm{AB}} \sqrt{\frac{4 \mathcal{D}_\mathrm{AB} t}{\pi}}.
    \label{eq:ConcentrationUnitMassFluxDiffusion}
\end{equation}
This formulation is time dependent; for the analytical solution of the diffusion problem, please refer to \cite{Personnettaz2019} and \cite{Herreman2020a}.
If solutal convection takes place, the variation across the layer is even lower and we can compute the latter a posteriori by numerical simulations.

The temperature differences within each layer can be estimated applying a 1D pure conduction model, like the ones presented by Köllner et al.~\cite{Koellner2017} and Personnettaz et al.~\cite{Personnettaz2018a}.
Here we use a simplified model based on the assumption that the thermal resistances ($\frac{\Delta h^\mathrm{(i)}}{k^\mathrm{(i)}}$) of the two liquid metal electrodes are comparable.
In this condition, the maximum of the temperature is situated at the mid height
of the electrolyte, where we expect no vertical heat flux.
Therefore, we can split the domain in two in order to decouple the electrodes.
The temperature jump across the full height of the $i$-th electrode can be expressed as
\begin{equation}
	\Delta \Theta^{(i)} =  \frac{\rho_\mathrm{el}^\mathrm{(E)} j^2 \Delta h^\mathrm{(E)}}{2}
	\frac{\Delta h^{(i)}}{k^{(i)}}  \ \mathrm{for} \ i = \mathrm{N, P}.
    \label{eq:TemperatureUnitElectrode}
\end{equation}
If additional heat fluxes are present at the interfaces, due to electrochemical
reactions, it is sufficient to add them to the first term.
In the electrolyte the maximum temperature variation spans across half of the electrolyte and can be estimated as
\begin{equation}
	\Delta \Theta^{\mathrm{(E)}} =  \frac{\rho_\mathrm{el}^\mathrm{(E)} j^2 (\Delta h^\mathrm{(E)})^2}{8 k^\mathrm{(E)}}.
    \label{eq:TemperatureUnitElectrolyte}
\end{equation}
This quantity was used as the temperature unit by Köllner et al.~\cite{Koellner2017}.
From these three simple formulas, we deduce that the temperature difference will
increase quadratically with the current density, but the concentration difference only linearly.
We can use Eq.~\ref{eq:TemperatureUnitElectrode} and
Eq.~\ref{eq:ConcentrationUnitMassFluxDiffusion} to estimate thermal and solutal Rayleigh
numbers in the liquid metal electrodes at a fixed current. \Tabref{t:numbers} gives an
overview on the estimated dimensionless numbers for different binary mixtures of the positive electrode.
In calculating these numbers, we assumed a layer height $\Delta h$ of \SI{1}{\centi\meter} for the
electrodes and the electrolyte, a current density of \SI{1}{\ampere\per\square\centi\meter} and
electrical resistivity of the salt of \SI{5e-3}{\ohm\per\meter}.
Most of the material has a Prandtl number substantially lower than 1. The diffusion of heat in the electrodes is very efficient.
It becomes clear that the solutal Rayleigh number is at least two orders of magnitude larger than the thermal Rayleigh number.
This suggests that solutal convection will be able to mix the thermally stably stratified positive electrode layer easily during charge.
On the other hand, flow induced in the positive electrode by viscous coupling with the electrolyte is expected to be strongly dampened by the compositionally stably stratified layer during discharge.
Two exceptions are electrodes made of Selenium or Tellurium. As both have a considerably lower thermal diffusivity compared to other liquid metals, we expect in such electrodes an interesting competition between heat and mass transfer.
\begin{table}
    \small
\caption{\label{t:numbers} Heat and mass transport dimensionless numbers for
    different combinations of active materials in the positive electrode. For
    the sources of the material properties, refer to \cite{Fazio2015,
	Iida2015}. The properties of the alloys (A-B) are computed at molar concentration of \SI{10}{\percent} of component A.
	Due to the scarcity of mixture properties we use blending rules, as described in \Sectref{s:material}.
	The mass transport diffusivity is estimated with the model proposed by Roy and Chhabra \cite{Roy1988}. $Ra_\mathrm{T,1}$,
	$Ra_{{\rho_A},1}$ are computed considering a $\Delta \Theta=\SI{1}{\kelvin}$, $\Delta\mathcal{C}=\SI{1}{\kilo\gram\per\cubic\meter}$, respectively.
    All values of Ra are given as absolute values. During discharge, $Ra_{{\rho_A}}$ is
    negative, \Eqref{eq:ConcentrationUnitMassFluxDiffusion} is evaluated at $t=$\SI{10}{\second}. Due to the stable temperature distribution, $Ra_\mathrm{T}$ is
    negative as well.}
\centering
\begin{tabular}{llllllllll}
\toprule
    name &   T &      Pr &      Sc &      Le &  $Ra_\mathrm{T,1}$ &
    $Ra_\mathrm{T}$  & $Ra_{{\rho_A},1}$  &   $Ra_{{\rho_A}}$  & $R_\rho$ \\
\midrule
Ca-Bi &\SI{550}{}&\SI{1.3e-02}{}&\SI{1.6e+01}{}&\SI{1.2e+03}{}& \SI{7e+02}{}&\SI{1e+03}{}& \SI{4e+06}{}&         \SI{3e+08}{}&        \SI{9e+00}{}\\
Ca-Sb &\SI{700}{}&\SI{1.7e-02}{}&\SI{4.0e+01}{}&\SI{2.3e+03}{}& \SI{4e+02}{}&\SI{5e+02}{}& \SI{4e+06}{}&         \SI{4e+08}{}&        \SI{20}{}\\
 K-Hg &\SI{250}{}&\SI{3.6e-03}{}&\SI{8.6e+00}{}&\SI{2.4e+03}{}& \SI{1e+04}{}&\SI{2e+04}{}& \SI{1e+08}{}&         \SI{4e+10}{}&        \SI{20}{}\\
 K-Tl &\SI{250}{}&\SI{1.8e-02}{}&\SI{1.6e+02}{}&\SI{9.2e+03}{}& \SI{3e+02}{}&\SI{3e+02}{}& \SI{2e+07}{}&         \SI{8e+09}{}&        \SI{40}{}\\
Li-Bi &\SI{450}{}&\SI{1.3e-02}{}&\SI{1.9e+01}{}&\SI{1.5e+03}{}& \SI{1e+03}{}&\SI{2e+03}{}& \SI{2e+07}{}&         \SI{6e+08}{}&        \SI{1e+00}{}\\
Li-Cd &\SI{493}{}&\SI{9.2e-03}{}&\SI{6.8e+01}{}&\SI{7.4e+03}{}& \SI{3e+02}{}&\SI{1e+02}{}& \SI{2e+07}{}&         \SI{1e+09}{}&        \SI{7e+00}{}\\
Li-Pb &\SI{500}{}&\SI{1.5e-02}{}&\SI{2.1e+01}{}&\SI{1.5e+03}{}& \SI{7e+02}{}&\SI{9e+02}{}& \SI{1e+07}{}&         \SI{4e+08}{}&        \SI{1e+00}{}\\
Li-Sb &\SI{450}{}&\SI{2.4e-02}{}&\SI{6.9e+01}{}&\SI{2.8e+03}{}& \SI{3e+02}{}&\SI{3e+02}{}& \SI{1e+07}{}&         \SI{5e+08}{}&        \SI{2e+00}{}\\
Li-Se &\SI{375}{}&\SI{1.0e+02}{}&\SI{6.3e+03}{}&\SI{6.2e+01}{}& \SI{9e+02}{}&\SI{8e+04}{}& \SI{3e+05}{}&         \SI{2e+07}{}&        \SI{0.1}{}\\
Li-Sn &\SI{400}{}&\SI{9.8e-03}{}&\SI{8.4e+01}{}&\SI{8.5e+03}{}& \SI{3e+02}{}&\SI{2e+02}{}& \SI{4e+07}{}&         \SI{2e+09}{}&        \SI{4e+00}{}\\
Li-Te &\SI{475}{}&\SI{1.7e-01}{}&\SI{8.4e+01}{}&\SI{4.8e+02}{}& \SI{3e+03}{}&\SI{3e+04}{}& \SI{2e+07}{}&         \SI{9e+08}{}&        \SI{0.4}{}\\
Li-Zn &\SI{486}{}&\SI{2.8e-02}{}&\SI{6.3e+02}{}&\SI{2.2e+04}{}& \SI{2e+02}{}&\SI{8e+01}{}& \SI{5e+07}{}&         \SI{5e+09}{}&        \SI{10}{}\\
Mg-Sb &\SI{700}{}&\SI{1.7e-02}{}&\SI{3.0e+01}{}&\SI{1.8e+03}{}& \SI{5e+02}{}&\SI{6e+02}{}& \SI{3e+06}{}&         \SI{4e+08}{}&        \SI{20}{}\\
Na-Bi &\SI{550}{}&\SI{1.0e-02}{}&\SI{6.1e+01}{}&\SI{6.0e+03}{}& \SI{1e+03}{}&\SI{2e+03}{}& \SI{5e+07}{}&         \SI{9e+09}{}&        \SI{10}{}\\
Na-Hg &\SI{275}{}&\SI{3.7e-03}{}&\SI{6.4e+00}{}&\SI{1.7e+03}{}& \SI{1e+04}{}&\SI{2e+04}{}& \SI{1e+08}{}&         \SI{1e+10}{}&        \SI{10}{}\\
Na-Pb &\SI{575}{}&\SI{1.2e-02}{}&\SI{2.8e+01}{}&\SI{2.4e+03}{}& \SI{7e+02}{}&\SI{8e+02}{}& \SI{1e+07}{}&         \SI{2e+09}{}&        \SI{10}{}\\
Na-Sn &\SI{625}{}&\SI{6.2e-03}{}&\SI{2.1e+01}{}&\SI{3.4e+03}{}& \SI{3e+02}{}&\SI{2e+02}{}& \SI{1e+07}{}&         \SI{9e+08}{}&        \SI{10}{}\\
\bottomrule
\end{tabular}
\end{table}

\subsection{Onset of convection in the electrolyte and negative electrode}
The onset of thermal convection in the negative electrode and electrolyte depends on current density, material properties and layer thicknesses; it can roughly be estimated using critical Rayleigh numbers from stability analysis available in the literature.
The Rayleigh number in the negative electrode can be estimated in first approximation combining \Eqref{eq:Ra_T} and \Eqref{eq:TemperatureUnitElectrode}:
\begin{equation}
    Ra^\mathrm{(\mathrm{N})} \approx \frac{g\beta_\mathrm{T}^\mathrm{(N)} \rho_\mathrm{el}^\mathrm{(E)} j^2 (\Delta h^\mathrm{(N)})^4 (\Delta h^\mathrm{(E)})}{2 k^\mathrm{(N)} \mathcal{D}_\mathrm{T}^\mathrm{(N)}\nu^\mathrm{(N)}}.
\end{equation}
In negative electrode, the temperature profile is linear for pure conduction, see \Figref{f:setup}.
The top surface is rigid (wall) and at constant temperature, the bottom surface can be well approximated by a uniform flux from the electrolyte.
This configuration with rigid bottom was studied by Sparrow et
al.~\cite{Sparrow1964}. They found a critical Rayleigh number of
$Ra_\mathrm{cr} = 1296$.
Köllner et al.~\cite{Koellner2017} performed a linear stability analysis of the
three-layer system, and found for the negative electrode a critical Ra of
1290, when the electrolyte is one tenth of the height of the electrodes.

The electrolyte has a parabolic temperature profile due to the Joule heating, as shown in \Figref{f:setup}.
There, the Rayleigh number is well approximated assuming an unstable length of half the layer thickness and the temperature difference proposed in \Eqref{eq:TemperatureUnitElectrolyte}:
\begin{equation}
	Ra^\mathrm{(\mathrm{E})} \approx \frac{g\beta_\mathrm{T}^\mathrm{(E)}
    \rho_\mathrm{el}^\mathrm{(E)} j^2 (\Delta h^\mathrm{(E)})^5}{64
    k^\mathrm{(E)} \mathcal{D}_\mathrm{T}^\mathrm{(E)}\nu^\mathrm{(E)}} = Ra_\mathrm{H}.
    \label{eq:Ra_E}
\end{equation}
This formulation is free of parameters of the other layers and consistent with previous work on internally heated convection \cite{KulackiGoldstein:1972}.
Kulacki and Goldstein~\cite{kulacki1975hydrodynamic} provided a complete description of the stability of internally heated layers subject to different combinations of thermal and mechanical boundary conditions.
In our application, the electrolyte is mechanical and thermally coupled with good heat conductors (liquid metal electrodes).
The conductivities and thicknesses of the electrodes, as well as any localised cooling or heating play a role for the onset of thermal convection within the molten salt.
Due to the low and comparable thermal resistance of the two layers, we assume an idealised condition in which the two interfaces are assumed isothermal at the same temperature.
The critical Rayleigh numbers in this condition of rigid-rigid and free-free interfaces are respectively 583 and 266 \cite{kulacki1975hydrodynamic}.
In a similar arrangement, Sparrow et al. \cite{Sparrow1964} show that the critical Rayleigh number for rigid interfaces is bounded between 560 and 583 for sufficiently conductive boundaries.

\begin{figure}[hbt]
\centering
\includegraphics[width=\textwidth]{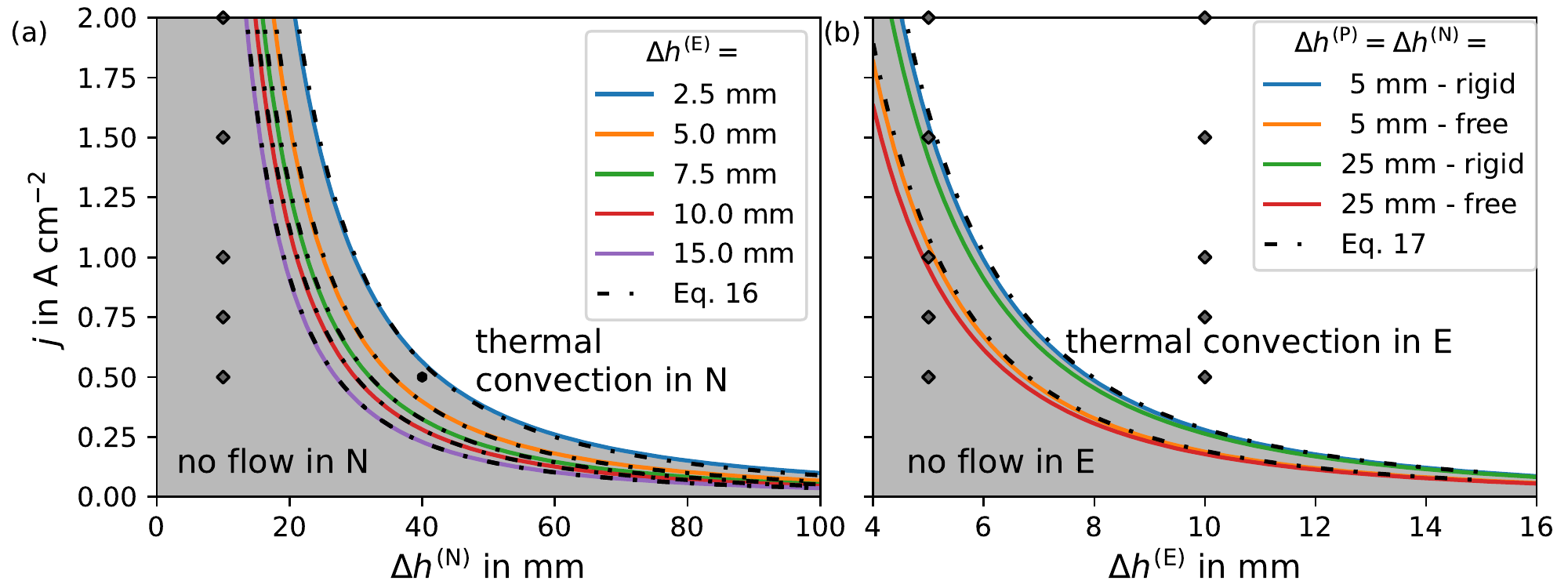}
\caption{Critical current density and layer heights for onset of
  convection in the negative electrode (a) and electrolyte (b). If not
  otherwise specified, the layers are 10\,mm thick; for the material
	properties, see Tab.~\ref{t:materials}. Continuous lines are computed with the pure diffusion model of \cite{Personnettaz2018a}.
	The dark grey rhombi correspond to the parameters used for the numerical
	studies \texttt{LARC}. The black hexagon in the left figure correspond to the \texttt{HARC} simulation.}
\label{f:onsetConvection}
\end{figure}

In order to get an impression of the relevance of thermal convection in realistic LMBs, we apply the critical Rayleigh numbers to the Li$||$Bi cell studied here.
The threshold current density, which corresponds to the critical Rayleigh number can be computed as function of the layer thickness and material properties using for the temperature difference an analytical model of pure conduction (e.g. \cite{Personnettaz2018a}), or by \Eqref{eq:TemperatureUnitElectrode} and \Eqref{eq:TemperatureUnitElectrolyte}.
The results, as illustrated in \Figref{f:onsetConvection}, show clearly that thermal convection in the negative electrode is not guaranteed. It will appear only in thick layers.
Using typical moderate current densities in the order of \SI{0.25}{\ampere\per\square\centi\meter} the electrolyte layer needs to be fairly thick as well for convection to set in, see \Figref{f:onsetConvection}b.
An efficient operation of the cell in terms of a low ohmic overpotential
($\eta_\Omega=\rho_\mathrm{el}j\Delta h^{\mathrm{(E)}}$) \cite{Kim2013a} requires thin electrolytes, which does not guarantee flow will appear in the latter.
The two simple formulae (\Eqref{eq:TemperatureUnitElectrode}, \Eqref{eq:TemperatureUnitElectrolyte}) provide a good approximation compared to the more complex analytical model of pure conduction and can be used to have a first prediction of temperature differences in the cells.
In order to proceed in our investigation we simulate heat and mass transport in LMB model.

\section{Numerical model}
The equations described in \Sectref{s:massTransport}, \Sectref{s:heatTransport} and \Sectref{s:fluidFlow} are discretized using the finite volume method, in the framework provided by the C++ library OpenFOAM \cite{jasak2020practical, Weller1998}.
The parent-child mesh strategy is applied to model the three-layer system with flat interfaces~\cite{Beale2016,Weber2017b}.
This means that besides of the global mesh taking care of the full domain, child meshes for each layer are generated.
Material properties can be seen as a field and they are consistently assigned to each mesh point.
A technique called mapping is used to exchange field data across the different domains and meshes.
For every time step, the system of equations is solved consecutively, with the following procedure:
(a) The advection-diffusion equation for the temperature, \Eqref{eq:AdvectionDiffusionTemperature}, is first solved on the parent mesh (full domain).
Continuity of temperature and heat flux at the interfaces are automatically enforced by this equation.
(b) In the second step, the mass transport equation, \Eqref{eq:AdvectionDiffusionMass}, is solved only in the positive electrode (child mesh).
The interface between the positive electrode and the electrolyte is treated as a boundary for this equation.
(c) Following these two steps, the new density for the buoyancy force is computed and mapped to each domain.
(d) Finally, the Navier-Stokes equations are solved sequentially in the negative electrode, the electrolyte and the positive electrode.
In order to ensure the mechanical coupling at the common interfaces, appropriate velocity interface conditions are required, as described in \Sectref{ss:layerCoupling}.
Knowing that that the mechanical coupling is performed in an explicit way (as
boundary conditions), the Navier-Stokes equations need to be solved several times in each phase until reaching convergence.
\subsection{Discretisation of the mechanical layer coupling}
\label{ss:layerCoupling}
In the OpenFOAM framework, the mechanical layer coupling must be ensured by ad hoc interface conditions. The latter we base here on a simplified version of Eq. 29 by Tukovic and Jasak \cite{tukovic2012moving}, neglecting the effects of a moving mesh and of surface tension. The exact derivation is the following:
the interface conditions for the velocity of a fluid 1 and 2 at a coupled interface read
\begin{align}
\bi u_\mathrm{f1} &= \bi u_\mathrm{f2},\\
\bi u\cdot \bi n &= 0,\\
\mu_1\nabla \bi u_\mathrm{t1}\cdot\bi n_1 &= \mu_2\nabla\bi u_\mathrm{t2}\cdot\bi n_2,
\end{align}
with $\bi u_\mathrm{f}$ denoting the velocity at the face between fluid 1 and 2,
$\bi u_\mathrm{t}$ the tangential velocity in the cell centre,
$\bi u_\mathrm{tf}$ the tangential velocity at the face,
$\bi n$ the face normal vector and $\mu$ the dynamic viscosity.
By combining the last two boundary conditions, we find
\begin{equation}
\mu_1\frac{\bi u_\mathrm{t1} - \bi u_\mathrm{tf}}{\delta_1} = \mu_2\frac{\bi u_\mathrm{tf}
  - \bi u_\mathrm{t2}}{\delta_2},
\end{equation}
with $\delta$ denoting the distance between cell centre and face.
This leads to
\begin{equation}
\bi u_\mathrm{tf} = w\cdot\bi u_\mathrm{t1} + (1-w)\cdot\bi u_\mathrm{t2}
\end{equation}
with the weighting factor
\begin{equation}
w = \frac{\delta_2\cdot \mu_1}{\delta_1 \mu_2+\delta_2 \mu_1}.
\end{equation}
As a tangential velocity can be defined as
\begin{equation}
\bi u_\mathrm{t} =\bi u\cdot(\bi I-\bi n\bi n),
\end{equation}
the boundary condition for the velocity at the interface might be written as
\begin{equation}
\bi u_\mathrm{tf} = w\cdot\bi u_{1}\cdot(\bi I-\bi n\bi n) + (1-w)\cdot\bi
u_{2}\cdot(\bi I-\bi n\bi n) = (\bi I-\bi n\bi n)\cdot(w\cdot\bi u_{1}
+ (1-w)\cdot\bi u_{2}),
\end{equation}
in which $\bi I$ denotes the identity matrix.

\section{Material properties and geometry}
\label{s:material}
\begin{figure}[hbt]
\centering
\includegraphics[width=0.8\textwidth]{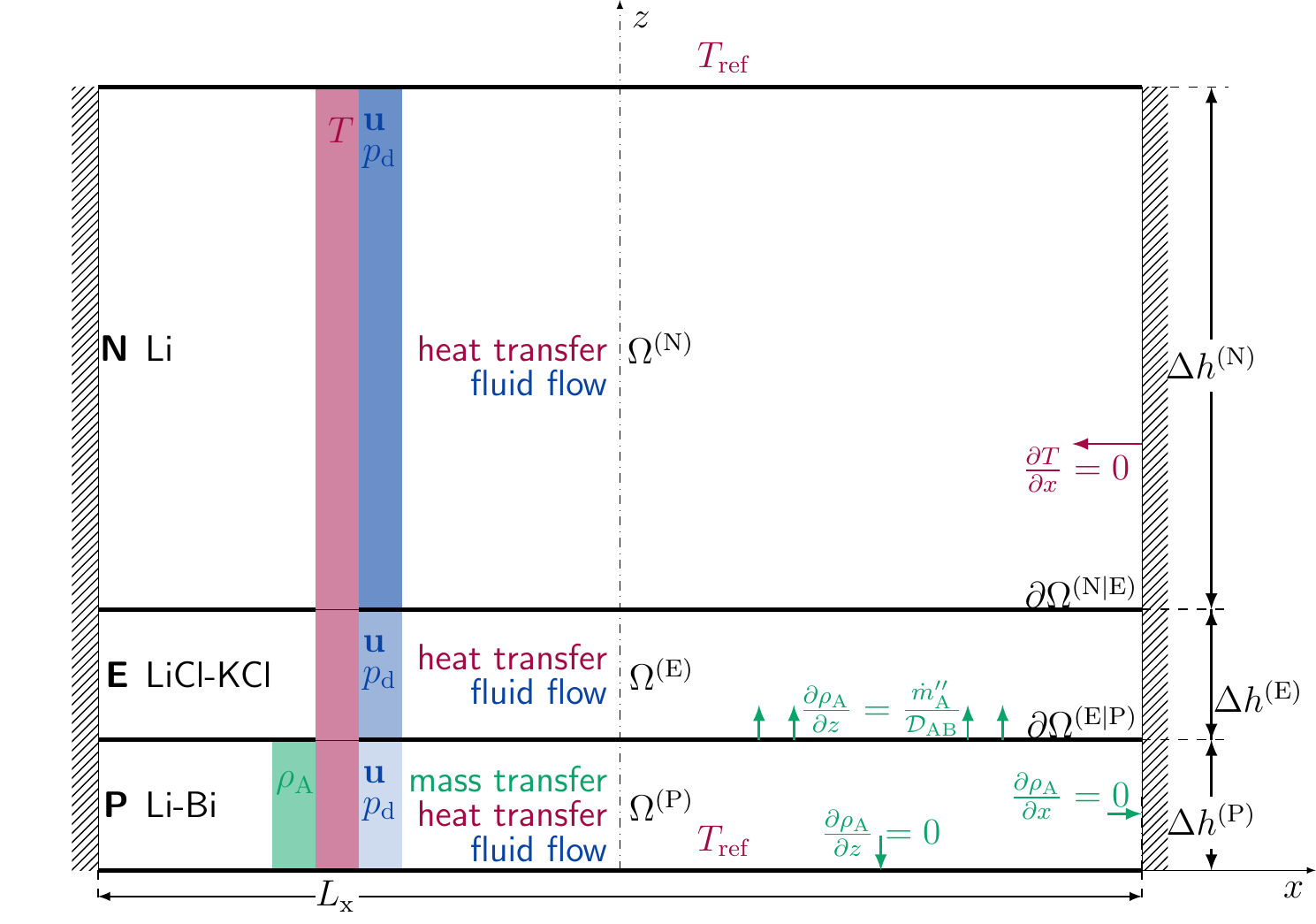}
\caption{Sketch of the 2D liquid metal battery model with temperature and concentration boundary conditions.
	The temperature is solved on the parent mesh (full domain), and the other equations on the child meshes (single regions).}
\label{f:geometry}
\end{figure}

We focus in the following on the well-investigated Li$||$Bi LMB at the reference temperature $T_\mathrm{ref}$ of \SI{450}{\celsius} in order to facilitate comparison of our results with previous literature \cite{Personnettaz2018a}.
Considering the excellent availability of material properties, we use a eutectic LiCl-KCl electrolyte, although KCl is known to be unstable in contact with Li at higher temperature \cite{Foster1964}.
Tab.~\ref{t:materials} gives an overview on the material properties of the three liquids.
Most of the properties of the Li-Bi alloy are not exactly known.
Therefore, the alloy density and solutal expansion coefficient have been estimated using Vegard's law as suggested by Fazio et al.~\cite{Fazio2015}.
We assume a reference molar fraction of $x_\mathrm{Li}=$~\SI{20}{\percent}, which corresponds to a Li-mass concentration of \SI{70.62}{\kilo\gram\per\cubic\meter}.
While the specific heat capacity is mass-weighted between Li and Bi according to the Neummann-Kopp's law~\cite{Iida2015}, the kinematic viscosity is molar weighted \cite{Personnettaz2018a}.
Thermal and electrical conductivity of the alloy are assumed to be equal to the one of pure Bi at the reference temperature.
\begin{table}[h]
	\caption{Thermodynamic and transport properties for a Li$||$Bi LMB at $T_\mathrm{ref} =\SI{450}{\celsius}$. The
	alloy properties have been calculated for a Li-molar fraction of \SI{20}{\percent}. $c_p$ is the specific heat capacity, which is used to compute the thermal diffusivity as $\mathcal{D}_\mathrm{T}=k(\rho_\mathrm{ref}c_p)^{-1}$.}\label{t:materials}
\centering
\begin{tabular}{llrrrr}
    \toprule
    \multicolumn{1}{c}{quantity} & \multicolumn{1}{c}{unit}             & \multicolumn{1}{c}{N - Li} & \multicolumn{1}{c}{E - LiCl-KCl} & \multicolumn{1}{c}{P - Li-Bi} &            \multicolumn{1}{c}{source} \\ \midrule
    $\rho_\mathrm{ref}$          & \si{\kilo\gram\per\cubic\meter}      &             \SI{4.911e2}{} &                   \SI{1.648e3}{} &                \SI{8.576e3}{} & \cite{Ohse1985,Raseman1960,Fazio2015} \\
    $c_p$                        & \si{\joule\per\kilo\gram\per\kelvin} &             \SI{4.237e3}{} &                   \SI{1.330e3}{} &                  \SI{1.7e2}{} & \cite{Ohse1985,Raseman1960,Fazio2015} \\
    $k$                          & \si{\watt\per\meter\per\kelvin}      &             \SI{5.218e1}{} &                  \SI{6.904e-1}{} &                \SI{1.421e1}{} &    \cite{Ohse1985,Janz1979,Fazio2015} \\ 
    $\rho_\mathrm{el}$           & \si{\ohm\per\meter}                  &             \SI{3.27e-7}{} &                 \SI{6.358e-03}{} &                \SI{1.39e-6}{} & \cite{Ohse1985,Raseman1960,Fazio2015} \\
    $\mathcal{D}_\mathrm{T}$     & \si{\square\meter\per\second}        &             \SI{2.51e-5}{} &                   \SI{3.15e-7}{} &                \SI{9.75e-6}{} &                                       \\
    $Pr$                         &                                      &             \SI{2.84e-2}{} &                      \SI{6.29}{} &                \SI{1.34e-2}{} &                                       \\ \midrule
    $\nu$                        & \si{\square\meter\per\second}        &             \SI{7.13e-7}{} &                 \SI{1.983e-06}{} &               \SI{1.304e-7}{} & \cite{Ohse1985,Raseman1960,Fazio2015} \\
    $\beta_\mathrm{T}$           & \si{\per\kelvin}                     &            \SI{1.923e-4}{} &                    \SI{3.2e-4}{} &               \SI{1.362e-4}{} &    \cite{Ohse1985,Janz1988,Fazio2015} \\ \midrule
    $\beta_{\rho_\mathrm{A}}$    & \si{\cubic\meter\per\kilo\gram}      &                            &                                  &               \SI{2.122e-3}{} &             \cite{Ohse1985,Fazio2015} \\
    $\mathcal{D}_\mathrm{AB}$    & \si{\square\meter\per\second}        &                            &                                  &                   \SI{7e-9}{} &               \cite{Personnettaz2019} \\
    $\mathcal{M}_\mathrm{A}$     & \si{\kilo\gram\per\mole}             &             \SI{6.94e-3}{} &                                  &                               &               \cite{linstrom2018nist} \\
    $Sc$                         &                                      &                            &                                  &                   \SI{18.6}{} & \\ \bottomrule
\end{tabular}
\end{table}
In order to resolve the flimsy plumes of solutal convection sufficiently -- and
taking into account the performance limitations of the finite volume solver --
we confine our study to two-dimensional simulations.
The geometry resembles the one shown in \Figref{f:geometry}.\\
The thickness of the electrolyte is bounded by safety and efficiency reasons (see \cite{Personnettaz2018a} details); here we use 5 and \SI{10}{\milli\meter}.
The necessary negative and positive electrode layer heights are determined by the desired composition of the
positive electrode in the discharged state. Assuming that all Bi of the positive electrode can be converted into the intermetallic phase
Li$_3$Bi, a negative electrode height of about 20\,mm would be sufficient\label{sec:layer_heights}. An alternative
 approach is to discharge only in the liquid phase region
of the Li-Bi phase diagram ($x_{\text{Li}} \lesssim $\SI{40}{\percent} at
\SI{450}{\celsius}). This would result in a minimum negative electrode height of approximately~4\,mm
required to fully discharge a cell containing a 10\,mm thick positive electrode.
Here, we investigate two configurations both with a \SI{10}{\milli\meter} positive
electrode, referred to as \texttt{HARC} (high aspect ratio cell) and \texttt{LARC}
(low aspect ratio cell), respectively.
The \texttt{HARC} has a \SI{40}{\milli\meter} thick negative electrode, see \Figref{f:geometry}.
In this configuration, thermal convection is also expected in the negative electrode at a current density of \SI{0.5}{\ampere\per\square\centi\meter} (see the black hexagon in \Figref{f:onsetConvection}a).
The low aspect ratio cell \texttt{LARC} with \SI{10}{\milli\meter} thick negative electrode is used to study the effect of solutal convection on internally heated convection.
Geometrical dimensions and applied current densities of the different configurations studied are collected in \Tabref{t:geometry_current}.
\begin{table}[h!]
    \centering
    \caption{Simulations list, geometry and current density. HT: heat transfer,
    MT: mass transfer.}
    \begin{tabular}{cllllll}
\toprule
               name         & $L_{x}$           & $\Delta h^\mathrm{(P)}$ & $\Delta h^\mathrm{(E)}$ & $\Delta h^\mathrm{(N)}$ & $j$                                 & properties           \\
                            & \si{\milli\meter} & \si{\milli\meter}       & \si{\milli\meter}       & \si{\milli\meter}       & \si{\ampere\per\square\centi\meter} &                      \\ \hline
          \texttt{HT\_PS}   & 135               & 45                      & 10                      & 45                      & 0.5                                 & \cite{Koellner2017}  \\
         \texttt{MT\_SEM}   & 40                & 10                      & 10                      & 10                      & 0.5                                 & \Tabref{t:materials} \\
         \texttt{HARC}      & 80                & 10 & 10                      & 40                      & 0.5, -0.5                                 & \Tabref{t:materials} \\
    \texttt{LARC\_H10}   & 80                & 10                      & 5                       & 10                      & 0.5, 0.75, 1.0, 1.5, 2.0            & \Tabref{t:materials} \\
   \texttt{LARC\_H5}     & 80                & 10                      & 10                      & 10                      & 0.5, 0.75, 1.0, 1.5, 2.0            & \Tabref{t:materials} \\
\bottomrule
    \end{tabular}
	\label{t:geometry_current}
\end{table}
\section{Comparison with spectral solvers}
In the absence of a suitable comparative case for the full problem,
validation of our model is performed separately for thermal and solutal
convection, each by comparing the OpenFOAM simulations with results of
a pseudo-spectral and a Fourier-spectral-element code respectively.
\subsection{Thermal convection}
The simulation of thermal convection in a three-layer LMB done with a
pseudo-spectral code by Köllner et al.~was employed here as a reference.
This test case has already been used by Personnettaz at. \cite{Personnettaz2018a}.
The solver relies on a spatial discretization in Fourier modes on the horizontal
plane and Chebychev polynomials in the vertical direction \cite{Koellner2017, Koellner2014}.
The horizontal discretization requires periodic boundary
conditions at the lateral walls.
We enforce the same boundary conditions also in our OpenFOAM simulation to allow for
comparison.

\begin{figure}[hbt]
\centering
\includegraphics[width=\textwidth]{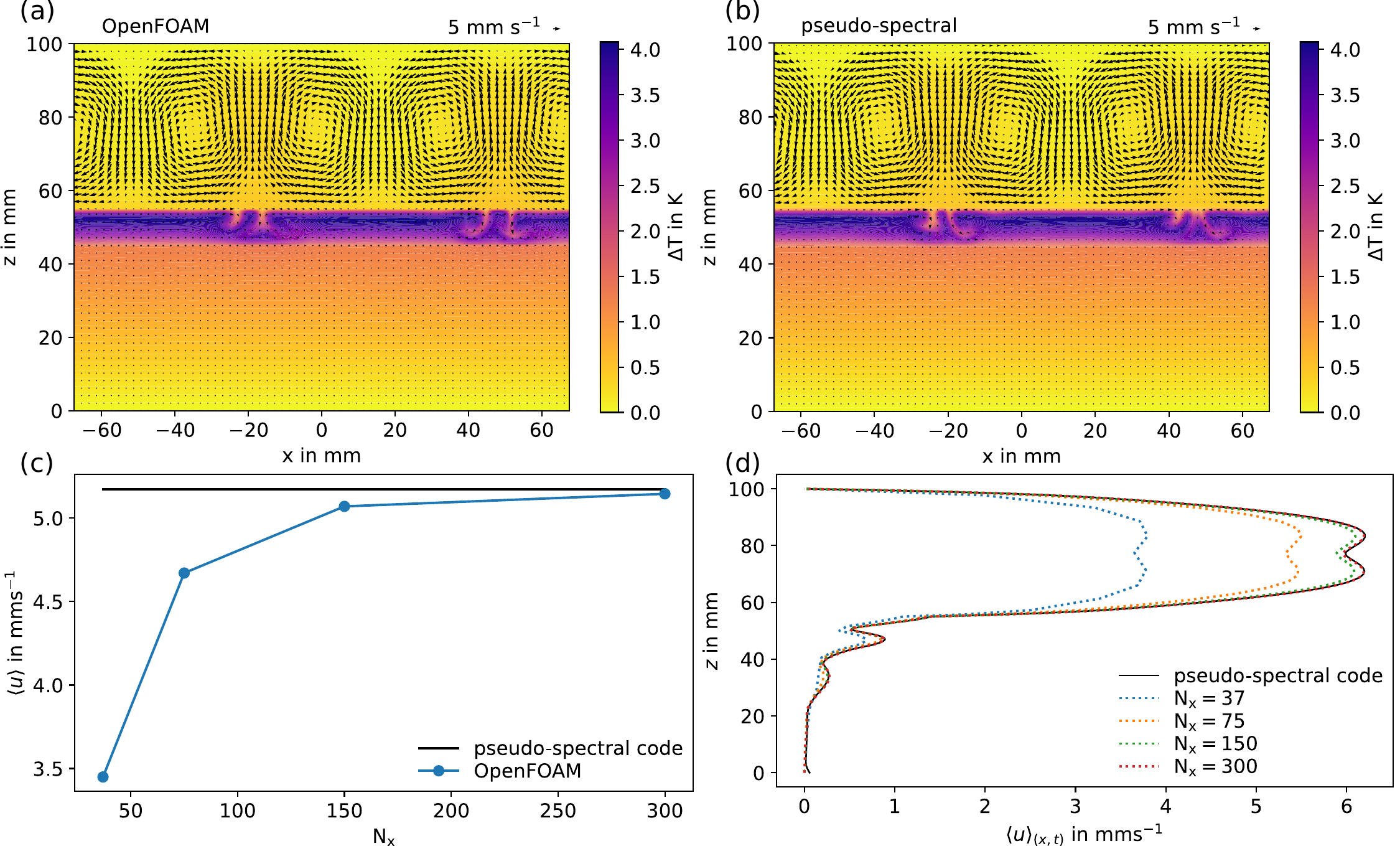}
\caption{Validation of thermal convection: (a) velocity and
  temperature distribution of OpenFOAM, (b) the same for the
	pseudo-spectral code at t = \SI{998}{\second}. (c) Volume averaged velocity
    and (d) space and time-averaged vertical velocity
    profile as a function of the lateral mesh
    resolution. Simulation \texttt{HT\_PS}.}
\label{f:validation:thermal}
\end{figure}

The same cell as the one studied by Köllner et al.~\cite{Koellner2017} is used here:
a Li$||$Pb-Bi LMB operating at \SI{500}{\celsius}; for the material properties, see \cite{Koellner2017}.
The geometry and current density are collected in \Tabref{t:geometry_current},
with the simulation being denoted as \texttt{HT\_PS}.
The velocity and temperature distributions, illustrated in \Figref{f:validation:thermal}a-b,
shows a clear resemblance between both numerical codes. By averaging
the velocity spatially over the x-coordinate, and thereafter over a
time of 50\,s, we obtain the vertical velocity profile as shown in
\Figref{f:validation:thermal}c. Very similar as in a recent
publication \cite{Personnettaz2018a}, we find that a mesh resolution
of ~300 control volumes over the lateral size yields a converged
velocity profile. Taking the average velocity in steady state condition (over space, and
50\,s again) we basically obtain the same result -- see \Figref{f:validation:thermal}d.

\subsection{Solutal convection}
\begin{figure}[hbt]
\centering
\includegraphics[width=\textwidth]{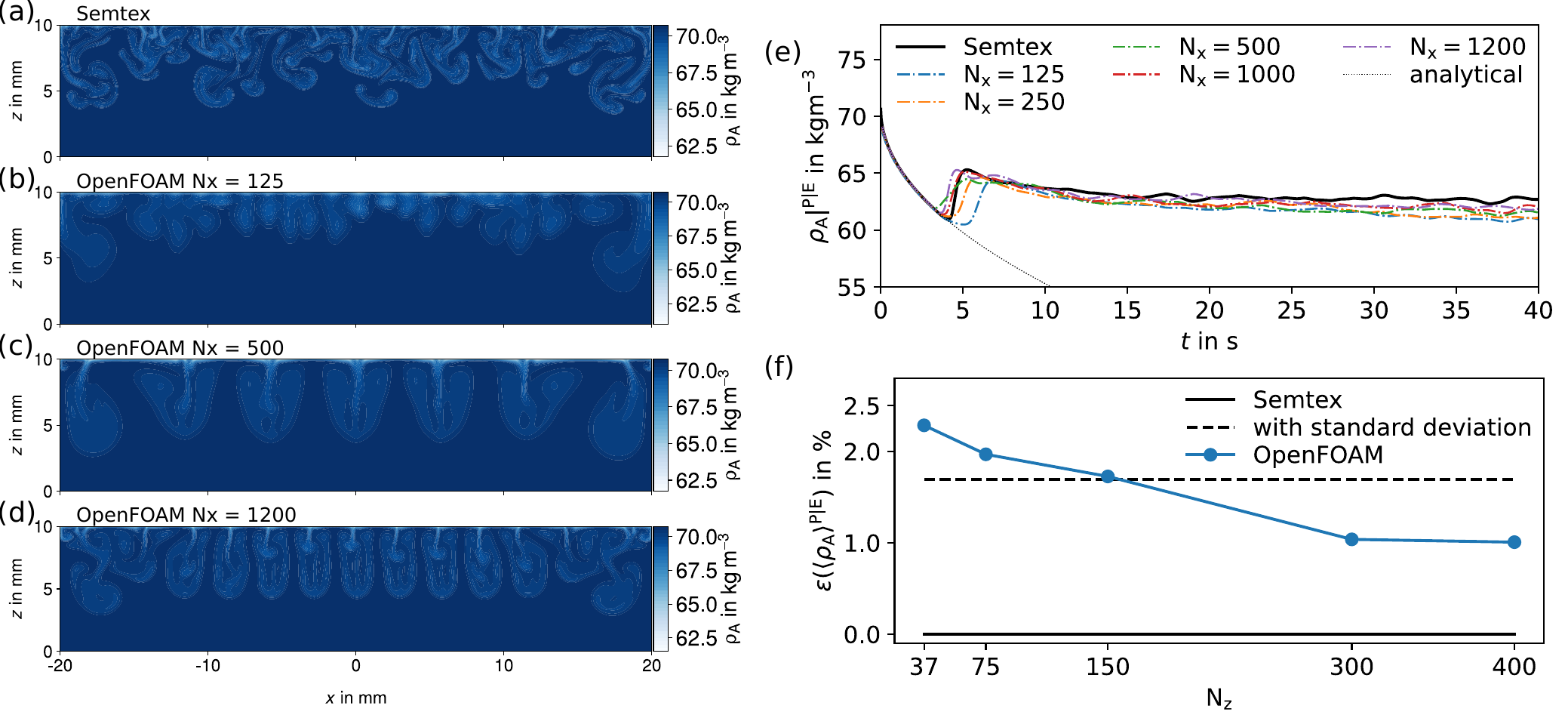}
\caption{Comparison of solutal convection: (a) concentration distribution obtained
	with Semtex and (b-d) OpenFOAM at different grid resolutions at $t=\SI{6.5}{\second}$, (e) time evolution of the mean mass concentration of Li at the
    interface $\partial\Omega^\mathrm{(P|E)}$ and (f) error of the Li interface concentration compared to Semtex, if averaged over the last \SI{15}{\second}.}
\label{f:validation:solutal}
\end{figure}

As a test case for solutal convection, we model the positive electrode of a Li$||$Bi LMB, as illustrated in \Figref{f:validation:solutal}.
The dimensions are given  in \Tabref{t:geometry_current} (test case \texttt{MT\_SEM}); the applied current density is \SI{0.5}{\ampere\per\square\centi\meter}.
The comparison is performed against a 2D simulation in Cartesian coordinates done with the Fourier-spectral element code Semtex \cite{Blackburn2019}.
In this framework, a grid independent solution was reached using 32 elements in vertical direction with 12x12 polynomials per element.
In order to perform the comparison, the boundary condition at the positive electrode/electrolyte interface is set to no-slip in both solvers.
The required grid resolution for solutal convection is substantially higher than for thermal convection due to the lower diffusivity.
Especially the plume number and their width depend crucially on the mesh resolution, as can be observed in \Figref{f:validation:solutal}a-d.
Fortunately, integral quantities, as the average mass concentration at the electrolyte interface, are much less sensitive to the grid resolution.
This means, even if the plumes structure is not perfectly resolved, the positive electrode is well mixed in any way.
Therefore, the Li interface-concentration results to be similar even for coarser meshes -- as illustrated in \Figref{f:validation:solutal}e-f.

Considering that within each time step, the equation system needs to be solved several times due to
the explicit interface coupling, it is extremely challenging to obtain the high
accuracy of the spectral codes with OpenFOAM in reasonable computational time.
Furthermore, solutal convection in the regime studied presents a strong
dependence on initial condition and level of numerical noise.
In conclusion, we select a mesh able to capture the main flow structures and
as well as integral quantities.
The cell is meshed with at least 600 control volumes in the horizontal direction and
200 over the positive electrode's vertical extension.
The boundaries, the positive electrode area as well as the interfaces
between the fluids are strongly refined.

\section{Results \& discussion}
We first present results of pure solutal convection in the three-layer system.
Then, we move to the interaction of heat and mass transfer.

\subsection{Solutal convection}
\label{s:solutalConvection}
\begin{figure}[hbt!]
\centering
\includegraphics[width=0.7\textwidth]{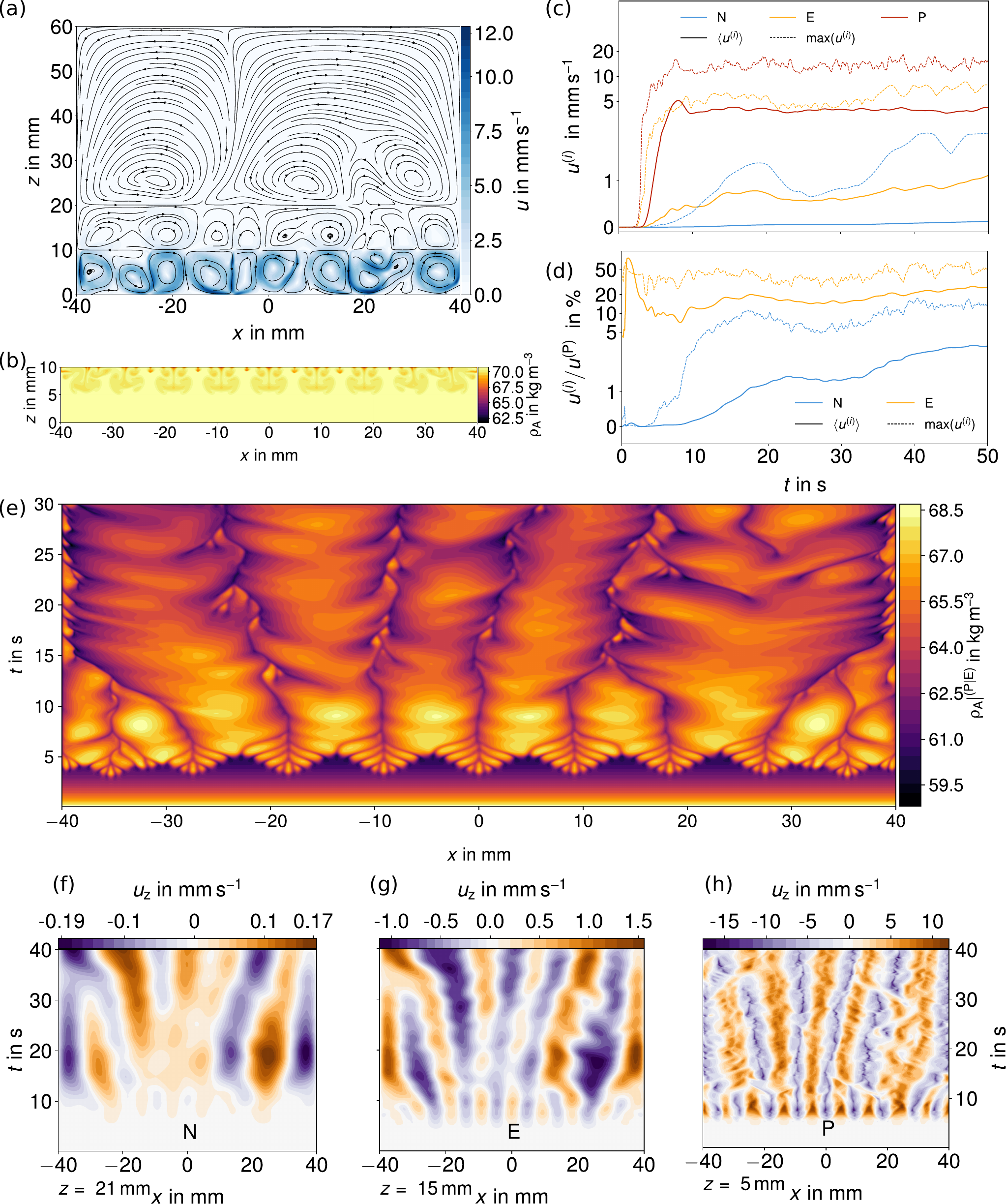}
\caption{Flow structure and magnitude at 30\,s (a), mass concentration
    distribution at 5.5\,s (b), maximum and volume averaged velocity (c), relative
  velocity compared to the one in the positive electrode (d),
  mass concentration at the positive electrode/electrolyte interface (e) and
  vertical velocity over a horizontal line in the negative electrode (f),
  electrolyte (g) and positive electrode (h). Simulation \texttt{HARC}.}
\label{f:solutal}
\end{figure}

We first present solutal convection alone in the \texttt{HARC} cell -- for the dimensions, see \Tabref{t:geometry_current}.
As reported before, the formation of the initial plumes involves three distinct phases: first, we observe a quasi-stable state,
where a concentration boundary layer builds up.
This layer is well described by the pure diffusion solution, see \Figref{f:solutal}.
Shortly after, the boundary layer becomes unstable, and finally small
plumes break rapidly out of the layer as shown in
\Figref{f:solutal}b \cite{goldstein1995onset}.
This complete process takes only 4\,s in our conditions.
As often observed before, the spatial wavelength increases
then with time due to merging of several plumes into larger convection
cells \cite{Philippi2019,Backhaus2011} -- this happens in our case within only 3
seconds (\Figref{f:solutal}e). After less than 10\,s, the complete
positive electrode is already well mixed. The secondary vortices, which develop
due to viscous coupling in the electrolyte and negative electrode layer turn as
expected in opposite direction  (\Figref{f:solutal}a)
\cite{prakash1997flow}. The mean averaged flow velocity in the positive electrode
reaches 5\,\si{\milli\meter\per\second}, in the electrolyte 1\,\si{\milli\meter\per\second} and in the negative electrode only
0.1\,\si{\milli\meter\per\second} (\Figref{f:solutal}c). Figures \ref{f:solutal}f-h illustrate
the vertical component of the velocity over time along a horizontal
line in the middle of each layer. Obviously, the number of convection
cells is very similar within the three phases. Flow starts first in
the positive electrode, slightly later in the electrolyte and finally in the
negative electrode. Momentum is effectively transferred across the layers.

The most important conclusions of our example are that solutal
convection needs only few seconds to drive strong flow in the positive electrode
and that it mixes the electrolyte layer efficiently as well. There,
the velocities reach 20\% of the mean averaged velocity of the positive electrode.
\subsection{Thermal convection and mass transport}
The effect of thermal convection on mixing and cell efficiency has
been discussed controversially in the past. Kelley \& Sadoway
\cite{Kelley2014} as well as Beltrán \cite{2016} suggested that bottom
heating might lead to mm-scale velocities in the positive electrode. Shen \&
Zikanov \cite{Shen2015} found that even the flow in the electrolyte might induce
velocities in the order of 0.3\,\si{\milli\meter\per\second} in the positive electrode by viscous
coupling in small LMBs; they further predicted that convection will be
much faster in large cells. Moreover, it has been shown that
electrochemical heating might even lead to an unstable temperature profile
in the positive electrode \cite{Personnettaz2018a} and that anti-convection might
appear, as well \cite{Koellner2017}. However, all these studies
neglected the interaction with mass transport that is always present in the
positive electrode during operation.

\begin{figure}[hbt]
\centering
\includegraphics[width=0.7\textwidth]{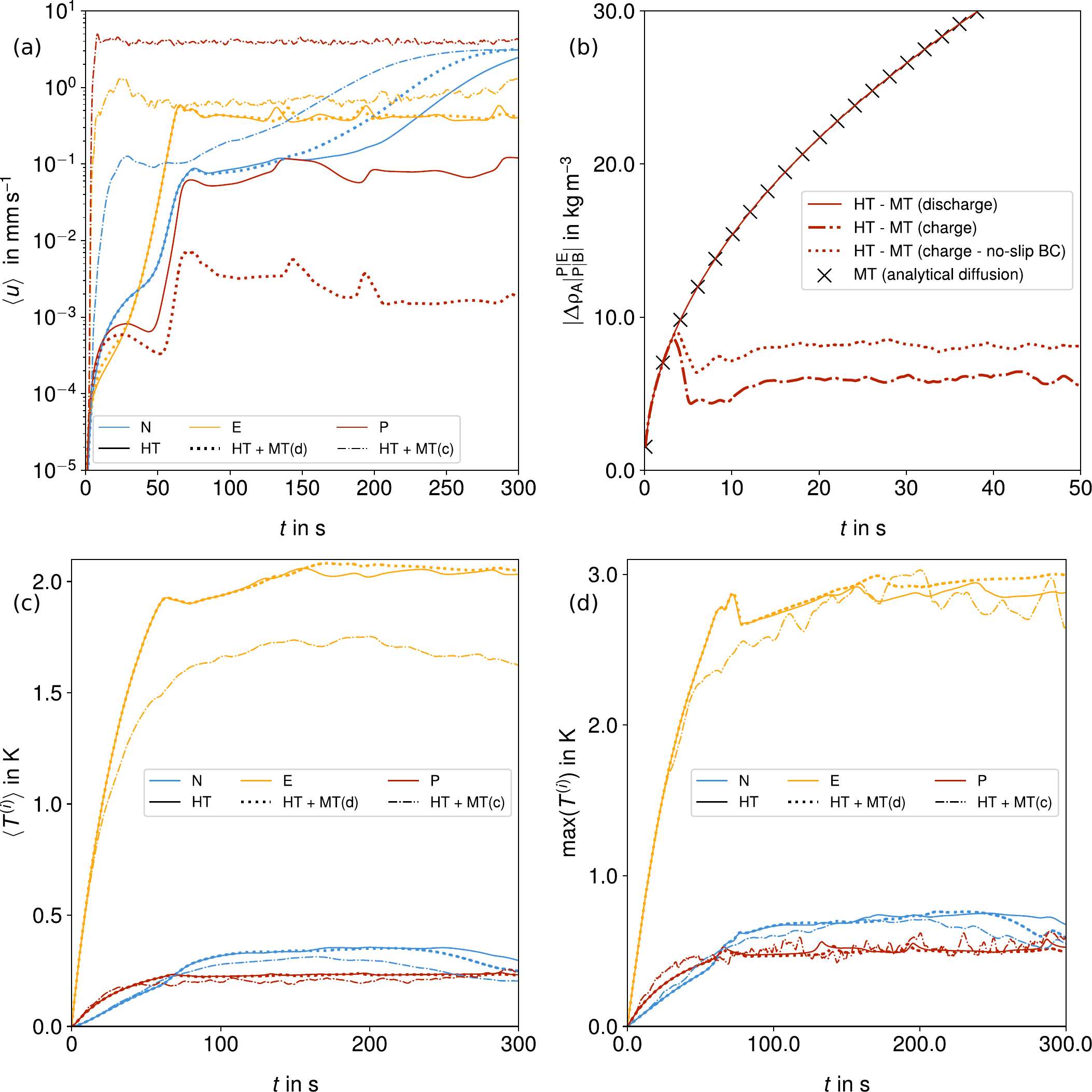}
	\caption{Volume averaged velocity (a), volume averaged temperature (c), maximum temperature (d) in the three layers
	in presence of pure thermal convection (HT), at discharge (HT + MT (d)) and charge (HT + MT (c)).
	Average concentration difference between bottom wall and positive electrode/electrolyte interface (b)
	during discharge, charge and charge with a rigid interface. \texttt{HARC} setup.}
\label{f:results:thermal}
\end{figure}

We focus our first investigations on the \texttt{HARC} setup, with the dimensions given in \Tabref{t:geometry_current}, starting with discharge and then studying charge.
As already discussed, during discharge, a stable density stratification forms
in the positive electrode due to the concentration, and to smaller extent due to the temperature distribution.
We study this phase using a current density of \SI{-0.5}{\ampere\per\square\centi\meter}.
The coupling with the electrolyte is the only source of motion in the positive electrode. As the latter is too weak, the flow is strongly damped by the presence of the stable stratification.
In \Figref{f:results:thermal}a we observe that the velocity in the positive electrode at discharge (red dots) is more than one order of magnitude less than in the case of pure thermal convection (red line).
The concentration distribution in the positive electrode is unperturbed and follows exactly the one predicted by an analytical solution, see \Figref{f:results:thermal}b \cite{Personnettaz2018a}.
The mean velocities in the other layers are not substantially affected, as shown in \Figref{f:results:thermal}a.
Finally, also the temperature distribution in the cell is similar to the one in
the presence of pure thermal convection, \Figref{f:results:thermal}c-d, as well.

When switching to charge, solutal flow occurs after only few seconds and establishes vigorous motion in the positive electrode as shown in \Figref{f:results:thermal}a.
In previous investigations, the electrolyte/positive electrode interface had been replaced by a no-slip boundary condition, when modelling the positive electrode alone \cite{Personnettaz2019,Herreman2021}.
Our more realistic multi-layer model shows that the concentration difference
within the positive electrode is now smaller compared to the single region
simulation, see \Figref{f:results:thermal}b. Obviously, replacing the electrolyte/positive electrode interface by a no-slip boundary condition attenuates flow and mixing too much, and overestimates therefore the mass transport overpotential.

It is also immediately mixing the electrolyte layer, flow there appears ten times
earlier than in the scenario with pure thermal convection.
The mean temperature in the electrolyte is reduced by the presence of solutal convection, as shown in \Figref{f:results:thermal}c.

Overall, solutal convection plays two roles: first it is efficiently transporting and
dissipating heat through the positive electrode, such as a forced convection source
at the positive electrode/electrolyte interface.
And, second, it is blending the electrolyte through the mechanical coupling.
In the \texttt{HARC} setup, the height of the negative electrode of \SI{40}{\milli\meter} leads to
a thermal Rayleigh number that exceeds the critical value for
onset of Rayleigh-B{\'e}nard convection even for the relatively low
Joule heat produced by a current density of \SI{0.5}{\ampere\per\square\centi\meter}
(\Figref{f:onsetConvection}, left).
The electrolyte layer then essentially acts as a heater at the lower interface
of the negative electrode. Such a configuration has been studied already
Goluskin (case RB3, \cite{Goluskin2016}).
The typical regular convection rolls of low-Ra Rayleigh-B{\'e}nard
convection develop, as shown in \Figref{f:Ekin}b.
The growth phase needs considerable time to develop, because first a sufficient temperature gradient has to be established.
However, solutal convection is able to introduce disturbances in the top layer, which can trigger an earlier onset.

Viscous coupling at the interfaces is important for the HARC geometry, as well. Especially for the
fully developed flow, the negative electrode takes then the driving role and
influences the flow in the electrolyte to a large extent. This
``dragging mode'' \cite{JohnsonNarayanan:1997} contributes to the
formation of relatively broad vortex cells in the electrolyte.

Computing the temporal mean value of the volume-averaged kinetic energy as \cite{Shen2015}
\begin{equation}
  \frac{E_{\text{kin}}}{\rho} = \frac{1}{V} \int_V |{\bi u}|^2 dV
  \label{eqn:Ekin}
\end{equation}
we find that the lowest energy can be observed in the electrolyte for the HARC case, see \Figref{f:Ekin}.
Since Joule heating is proportional to the square of the current density, but mass transfer
depends only linearly on $j$, we expect that the maximum velocities in the negative electrode exceed
those in the positive electrode for the HARC geometry if the current density exceeds $0.5$\,A/cm$^{2}$.

\subsection{Interaction of thermal and solutal convection in low aspect ratio cells}
\begin{figure}[t]
\centering
\includegraphics[width=0.7\textwidth]{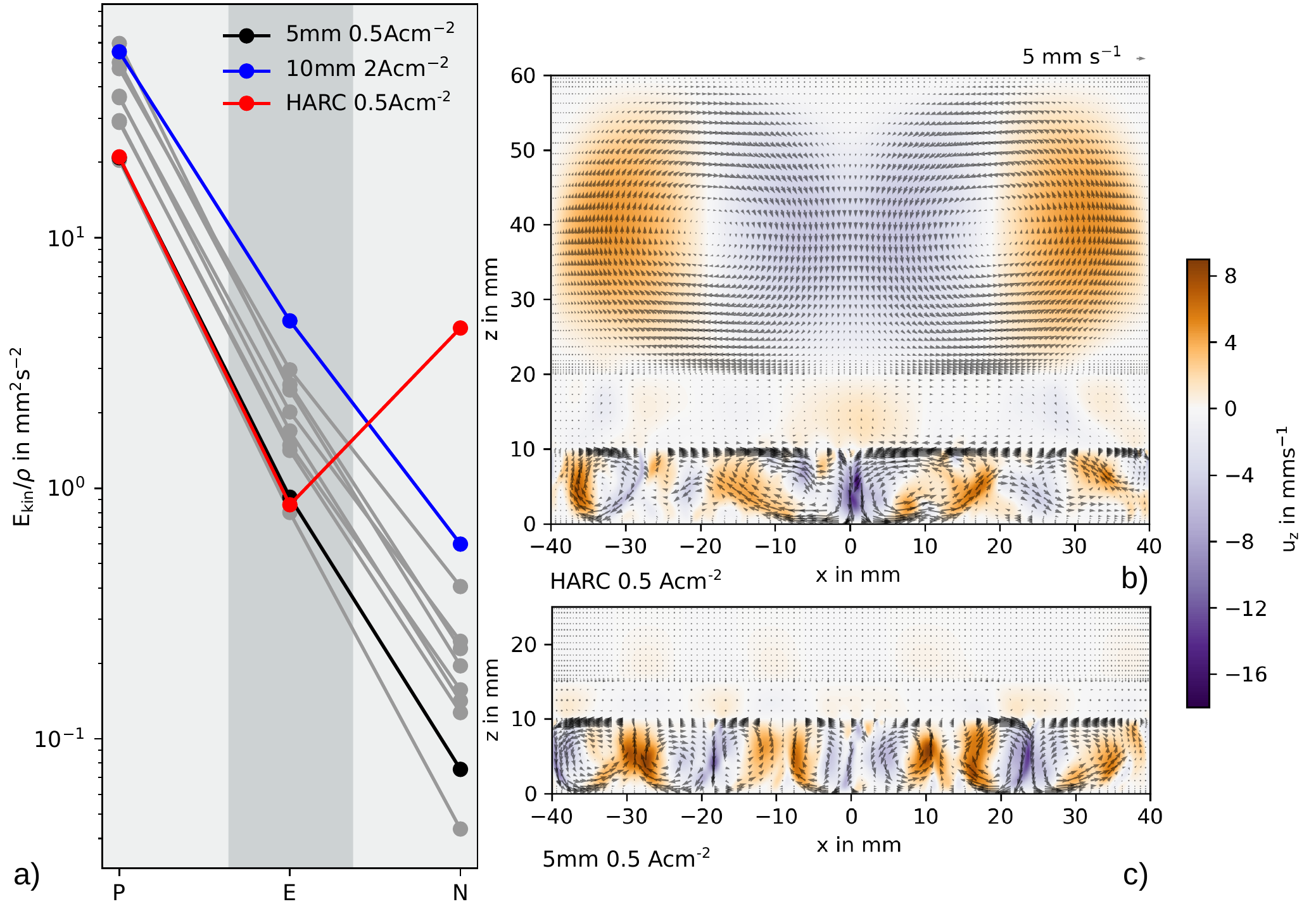}
\caption{Kinetic energy in the three layers of the cell for different
  current densities and aspect ratios (a). Snapshots of two flow
  fields for the cell with the high aspect ratio negative electrode (HARC, b) and for
	a LARC cell (c) under charge with \SI{0.5}{\ampere\per\square\centi\meter}.}
\label{f:Ekin}
\end{figure}

We limit the discussion in this section to the charging phase in vertically symmetric cells. The geometry is denoted
as ``low aspect ratio cell'' (LARC) in the text.
\Tabref{t:geometry_current} provides the dimensions and current density, which are also illustrated by grey rhombi in the right diagram
of \Figref{f:onsetConvection}.
The setup used here is much closer to real cells, as already discussed in \Sectref{sec:layer_heights}.
Looking at \Figref{f:Ekin}a it can be seen that the temporal mean value of the
volume-averaged kinetic energy decreases monotonically from the
positive electrode over the electrolyte to the negative electrode.

Withing all simulations, the solutal convection's flow velocity in the positive electrode exceeds
that in the other layers by at
least an order of magnitude in terms of kinetic energy.

In all cases, the layer coupling is dominated by viscous forces as already found by other authors \cite{Shen2015,Koellner2017} for
thermal convection in liquid metal batteries. The horizontal velocity
components near the interfaces go into the same direction on both
sides of the interface, and interface normal velocity components mirror
each other.

The total heat released in the electrolyte grows with the electrolyte
thickness and the current density. While the negative electrode's Rayleigh number is
smaller than the critical value for all current densities considered,
the electrolyte's Rayleigh number exceeds the critical value only for
$j=2$\,Acm$^{-2}$ and a 5\,mm thick electrolyte and for all current
densities, if the electrolyte thickness amounts to 10\,mm.

Thermally driven convection in the cell means classical
Rayleigh-B{\'e}nard convection in the negative electrode, but internally heated
convection in the electrolyte. The former is well investigated,
meticulously described in almost all details and still subject of
intense research; the latter received much less attention but is
nevertheless important for topics as diverse as mantle convection,
nuclear reactor engineering, and astrophysics. For a
relatively recent comprehensive review of both, see \cite{Goluskin2016}.

Solutal convection provides a momentum source to the electrolyte
via viscous coupling at the positive electrode/electrolyte interface.
\begin{figure}
  \centering
  \includegraphics[width=0.7\textwidth]{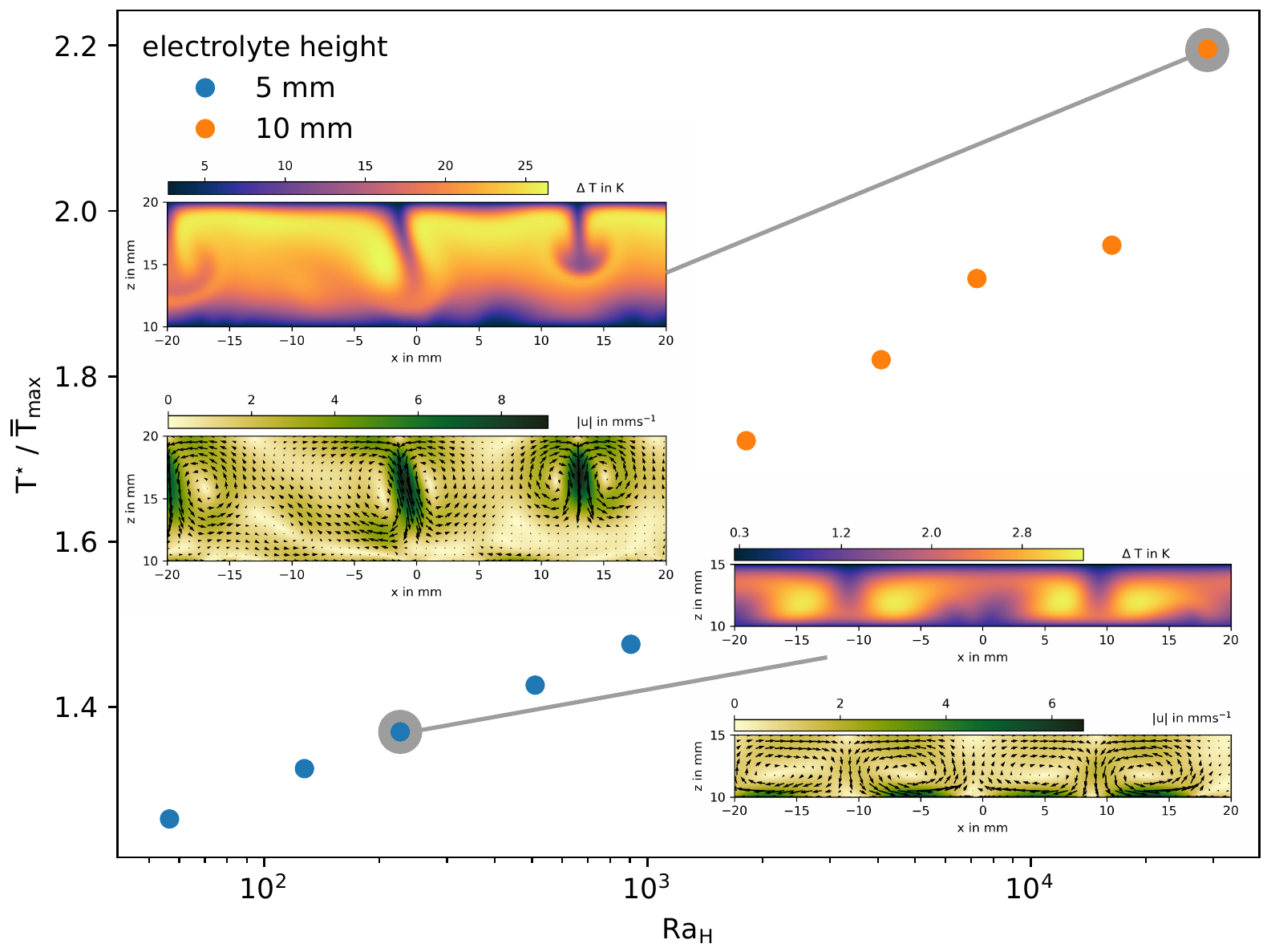}
  \caption{Convective cooling parameter
    $T^{\star}/\overline{T}_{\text{max}}$ as a function of Ra$_{\text{H}}$ for two
    different electrolyte heights. The two insets show snapshots of
    the temperature and velocity distribution of the fully developed
    flow in the electrolyte at the two parameter sets encircled in
    grey. In horizontal direction only the central part of the
	electrolyte is shown. \texttt{LARC} setup.}
  \label{f:convective_cooling_p}
\end{figure}
The influence of this momentum source on the flow in the electrolyte
is visible in the flow structure and the integral properties of the
thermal convection as can be seen in
\Figref{f:convective_cooling_p}. It shows the convective cooling
parameter $T^{\star}/\overline{T}_{\text{max}}$ as defined by Peckover
and Hutchinson \cite{PeckoverHutchinson:1974} vs.~the Rayleigh number
based on the Joule heat release and the electrolyte layer's half
height \cite{KulackiGoldstein:1972} described by \Eqref{eq:Ra_E}.
The convective cooling parameter is the ratio of the maximum
temperature $T^{\star}$ that would occur if heat transport were purely
conductive, to the maximum value of the (height dependent) temperature
profile $\overline{T}_{\text{max}}$ in the fully developed flow
averaged over time and in horizontal direction. The location of
$T^{\star}$ and $\overline{T}_{\text{max}}$ is marked in the
temperature profiles of \Figref{f:temperature_profiles_vs_j}. Even
for the smallest Ra$_{\text{H}}$ ($\text{Ra}_{\text{H}} = 56$), the cooling parameter
($T^{\star}/\overline{T}_{\text{max}} = 1.26$) exceeds
one. This signals that convection is present in the electrolyte
intensifying heat transfer. The lower inset in
\Figref{f:convective_cooling_p} ($\text{Ra}_{\text{H}} = 226,
T^{\star}/\overline{T}_{\text{max}}=1.37$) displays snapshots of the
instantaneous flow and temperature fields in the midsection ($-20\,\text{mm}
< x < 20\,\text{mm}$) of the electrolyte. Despite the Rayleigh number being
considerably lower than its critical value, four convection cells are
visible in the flow field. They are driven by viscous coupling at the
positive electrode/electrolyte interface ($z=\SI{10}{\milli\meter}$) where the highest velocities
can be observed. The temperature field is modulated by the influence
of convection and displays cooler regions where the flow transports
cold fluid from the boundaries into the bulk. Thus, the four
convection cells leave their footprints in the temperature field and
alter it considerably with respect to the purely conductive
case. While the velocities in the upper part of the electrolyte are
smaller than that in the lower part, the cells as well as their
thermal footprint occupy the entire electrolyte region.
%
%

\begin{figure}
  \centering
  \includegraphics[width=0.7\textwidth]{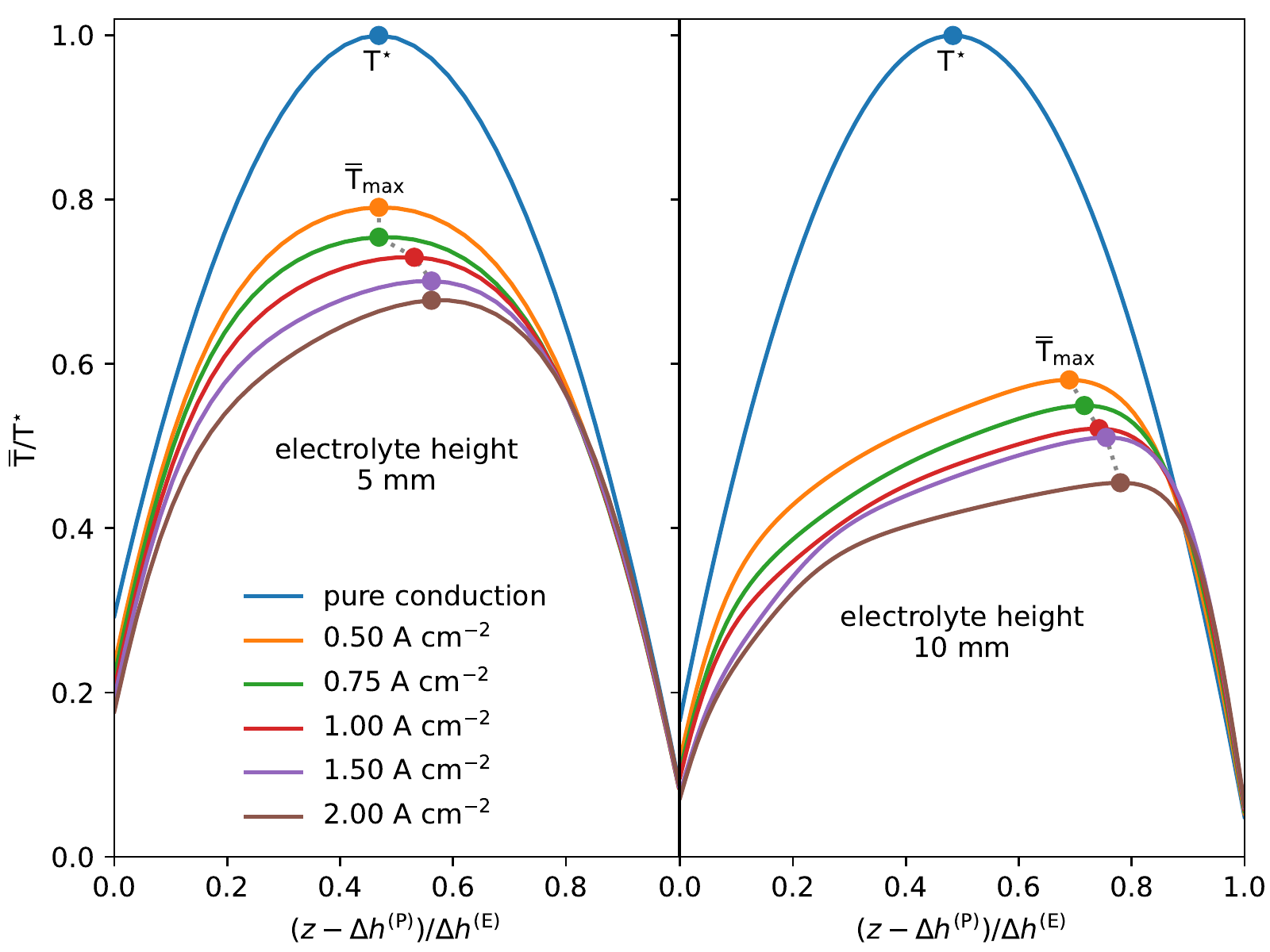}
  \caption{Horizontally averaged vertical temperature profiles in the
    electrolyte normalised with the maximum conductive temperature
    T$^{\star}$ for the 5\,mm and the 10\,mm thick electrolyte and
	different current densities. \texttt{LARC} setup.}
  \label{f:temperature_profiles_vs_j}
\end{figure}
The upper inset in \Figref{f:convective_cooling_p} ($\text{Ra}_{\text{H}}=28979,
T^{\star}/\overline{T}_{\text{max}}=2.2$) shows flow and temperature
fields of well developed internally heated convection. The flow
field looks now quite different and so does the temperature
field. While the influence of viscous coupling is still visible at the
positive electrode/electrolyte interface, the highest velocities now occur in the
plumes descending from the electrolyte/negative electrode interface because of the
unstable temperature and thereby density distribution. As
characteristic for internally heated convection with heat transfer
over both horizontal boundaries, convection is very intense in the
upper half of the layer, but damped in the lower one. This is a
consequence of the -- on average -- stable stratification in the lower
and the unstable stratification in the upper part of the layer. This
asymmetry results in penetrative convection (see, e.g.~\cite{Veronis:1963}) where fluid undergoes convective motion in the
upper part and penetrates into the lower stably stratified layer. Such
a situation leads to irregular motion even for Rayleigh numbers close
to the critical one as reported by a number of authors, e.g.,
\cite{JahnReineke:1974,RalphMcGreevyPeckover:1976,Tveitereid:1978,KulackiRichards:1985}
and as observed here as well.

\Figref{f:temperature_profiles_vs_j} displays the horizontally
averaged vertical temperature profiles normalised with the maximum
conductive temperature for the two electrolyte heights investigated.
As already discussed in connection with
\Figref{f:convective_cooling_p}, we observe convection for the
smallest Rayleigh number investigated (Ra$_{\text{H}} = 56$, corresponding to
$j=0.5$\,\si{\ampere\per\square\centi\meter} in the 5\,mm thick electrolyte). This results in a
moderate reduction of the maximum temperatures
$\overline{T}_{\text{max}}$ with respect to $T^{\star}$. However, the
shape of the temperature profiles is still relatively symmetric for
the 5\,mm thick electrolyte with somewhat larger deviations for
$j=$\SI{1.5}{\ampere\per\square\centi\meter} and 2\,\si{\ampere\per\square\centi\meter}. This changes for the 10\,mm thick
electrolyte with its much higher Rayleigh numbers due to the stronger
heat release in the thicker layer. The reduction of
$\overline{T}_{\text{max}}$ with respect to $T^{\star}$ is more
pronounced as is the asymmetry of the temperature profiles. The
location of $\overline{T}_{\text{max}}$ shifts to almost
$(z-\Delta h^{\mathrm{(P)}})/\Delta h^{\mathrm{(E)}}=0.8$ for
$j=$\SI{2.0}{\ampere\per\square\centi\meter}.  The increase in asymmetry with the heating rate in
the electrolyte is again a feature that is to be expected and known
from the literature, e.g., \cite{KulackiGoldstein:1972,
  PeckoverHutchinson:1974, Tveitereid:1978, KulackiRichards:1985}.

\begin{figure}
  \centering
  \includegraphics[width=0.7\textwidth]{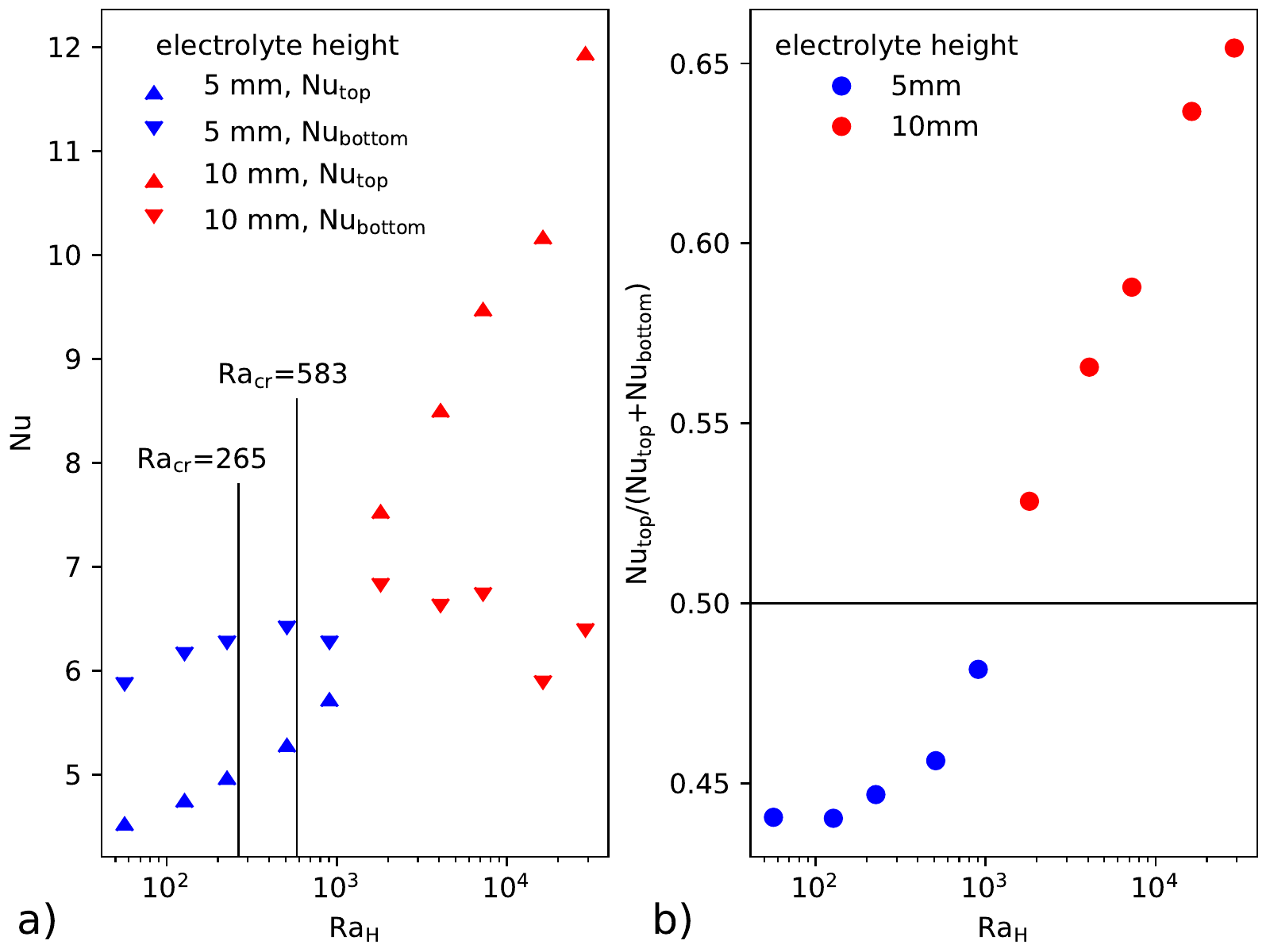}
  \caption{Nusselt numbers for the top and bottom interface of the
    electrolyte with the electrodes (left) and Nusselt number ratio
	(right) vs.~Rayleigh number for different electrolyte thicknesses. \texttt{LARC} setup.}
  \label{f:Nu_top_bottom}
\end{figure}
Corresponding to these changes in the mean temperature profiles, the
heat transfer conditions at the interfaces change. This is shown in
\Figref{f:Nu_top_bottom}. \Figref{f:Nu_top_bottom}a displays the
Nusselt numbers at the electrolyte/negative electrode interface (Nu$_{\text{top}}$)
and at the positive electrode/electrolyte interface  (Nu$_{\text{bottom}}$). The
Nusselt number definition is that used by Kulacki and Goldstein
\cite{KulackiGoldstein:1972}
\begin{equation}
\text{Nu} = \frac{\Delta h^{\mathrm{(E)}}
  |dT/dz|_{\text{interface}}}{T_{\text{max}} - T_{\text{interface}}},
  \label{eqn:Nusselt}
\end{equation}
with the electrolyte height $\Delta h^{\mathrm{(E)}}$, the temperature
gradient at the interface $|dT/dz|_{\text{interface}}$, the maximum
temperature in the layer $T_{\text{max}}$, and the temperature of the
interface $T_{\text{interface}}$. Internally heated convection
with equal temperatures at the horizontal boundaries features
typically higher Nusselt numbers on the top of the layer and lower
ones on the bottom, see, e.g.~\cite{KulackiGoldstein:1972}. This is in
contrast to what we observe for the 5\,mm thick electrolyte
layer. Here, Nu$_{\text{bottom}}$ exceeds Nu$_{\text{top}}$ for all
investigated Rayleigh numbers. Since viscous coupling at the lower
interface is the only source of convective motion for Ra$_{\text{H}} <
\text{Ra}_{\text{H, crit}}$, stronger convective heat transfer at the
lower interface is an expected result for these cases. But even for
the higher Ra in the 5\,mm thick electrolyte convective heat transport
due to viscous coupling outweighs that due to thermal convection.

The conditions are reversed for the 10\,mm thick electrolyte. As in
purely internally heated convection, Nu$_{\text{top}}$ is larger than
Nu$_{\text{bottom}}$ for all Ra$_{\text{H}}$ investigated. Thermal convection
clearly dominates momentum transfer by viscous coupling. This
transition can be observed as well by looking at the flux ratio
Nu$_{\text{top}}$/(Nu$_{\text{top}}$+Nu$_{\text{bottom}}$) in
\Figref{f:Nu_top_bottom}b. The flux ratio would be 0.5 for the
conductive case but lies below this value for the 5\,mm thick
electrolyte, i.e., more heat is transferred through the
positive electrode/electrolyte interface than through the electrolyte/negative electrode
interface. For the 10\,mm thick electrolyte, the flux ratio is larger
than 0.5 because more heat leaves the electrolyte through the top
interface than through the bottom one.

\section{Summary}
As liquid metal batteries (LMBs) operate as concentration cells,
mixing is extremely important. After discussing that concentration
overpotentials might appear in the positive electrode, we
briefly gave an overview on the various flow phenomena being present in
LMBs. We showed that the effects of solutal convection on the cell
voltage have already been observed 60 years ago. However, the source
of these effects was explained only very recently by Personnettaz et
al.~\cite{Personnettaz2019}.

With this background and motivation, we studied the interaction of
solutal and thermal convection in a Li$||$Bi LMB. A brief theoretical
analysis suggests that solutal convection will always be able to mix the
stable thermal stratification. On the other hand, the thermal flow is
not expected to be strong enough to mix the compositionally stably
stratified positive electrode. Using critical Rayleigh numbers from
literature, we predicted further that thermal convection will appear
only in very thick electrolyte layers and seldomly in the negative electrode.

In order to study the interplay of solutal and thermal convection in more detail, we developed a
numerical model with fixed interfaces, solving for the temperature and
flow field within the complete cell, and the concentration distribution in
the positive electrode. The validation, which was performed by comparison with
two (pseudo-)spectral codes, suggests that the mesh resolution for solutal
convection needs to be considerably greater than for thermal flow.

In a first step, we simulated solutal convection, which is present only in the
positive electrode during the charge phase. We found that it almost instantly produces a very
strong mixing effect in the cell. The velocities, which are induced in the electrolyte via
viscous coupling, reach still 20\% of the mean average velocity compared to the
positive electrode. Therefore, we deduced that compositional flow mixes both the
positive electrode and electrolyte. We further found that it is important
to model the mechanically coupled electrolyte/positive electrode interface.
Replacing the interface with a simplified no-slip boundary condition
attenuates flow and mixing and will therefore overestimate the mass
transport overvoltage \cite{Herreman2020a,Personnettaz2019}.

In a second step, we simulated thermal convection in discharge. As it appears
first in the electrolyte, it is able to mix this region well. However,
the viscous coupling is not strong enough to induce any relevant flow
in the positive electrode.

Finally, we studied thermal and solutal
convection together. We showed that thermal convection needs roughly
ten times longer to develop, compared to the solutal flow. The
layer coupling is always viscous. While thermal convection does not
develop in thin electrolyte layers, except for the highest
investigated current density, thick electrolyte layers are always dominated
by thermally driven flow. Accordingly, thin electrolyte layers feature
relatively regular convection cells during charge. These extend over
the full height and are driven by viscous coupling from the
positive electrode that is well mixed by solutal convection. In contrast, the
Rayleigh numbers of thick electrolyte layers are well above the critical
ones for all cases considered. Averaged temperature profiles show a
pronounced asymmetry typical for penetrative convection. Nusselt
numbers at the upper boundary of the electrolyte layer exceed that at
the lower boundary for thick electrolytes. The opposite holds true for
thin electrolytes, where viscous coupling drives the most intense flow
of the lower boundary and the cooling parameter is smaller than 0.5.

In summary, we would like to stress that solutal convection is a very
powerful effect. Although it appears only in the positive electrode at charge, it
is able to mix then the electrolyte via viscous coupling at the
interface. In contrast, thermal convection can appear only in
comparably thick electrolyte layers, but will then allow mixing them
during charge and discharge.

\section*{\small Acknowledgements}
This project has received funding from the European Union’s Horizon
2020 research and innovation programme under grant agreement No
963599, from the Deutsche Forschungsgemeinschaft (DFG, German
Research Foundation)  by  award  number  338560565 and in frame of the
Helmholtz - RSF Joint Research Group ``Magnetohydrodynamic
instabilities: Crucial relevance for large scale liquid metal
batteries and the sun-climate connection'', contract No. HRSF-0044 and
RSF-18-41-06201. The authors are grateful to T. Köllner for the pseudo-spectral
simulation of thermal convection in a three-layer LMB, used here for validation.
Fruitful discussions with S. Bénard, W. Herreman and
C. Nore are gratefully acknowledged.

\bibliography{extracted}

\begin{thebibliography}{10}
\expandafter\ifx\csname url\endcsname\relax
  \def\url#1{\texttt{#1}}\fi
\expandafter\ifx\csname urlprefix\endcsname\relax\def\urlprefix{URL }\fi
\expandafter\ifx\csname href\endcsname\relax
  \def\href#1#2{#2} \def\path#1{#1}\fi

\bibitem{Kim2013a}
H.~Kim, D.~A. Boysen, T.~Ouchi, D.~R. Sadoway, Calcium - bismuth electrodes for
  large - scale energy storage (liquid metal batteries), J. Power Sources 241
  (2013) 239--248.

\bibitem{Kelley2018}
D.~H. Kelley, T.~Weier, Fluid mechanics of liquid metal batteries, Appl. Mech.
  Rev. 70~(2) (2018) 020801.
\newblock \href {https://doi.org/10.1115/1.4038699}
  {\path{doi:10.1115/1.4038699}}.

\bibitem{Personnettaz2019}
P.~Personnettaz, S.~Landgraf, M.~Nimtz, N.~Weber, T.~Weier, Mass transport
  induced asymmetry in charge/discharge behavior of liquid metal batteries,
  Electrochem. Commun. 105 (2019) 106496.
\newblock \href {https://doi.org/10.1016/j.elecom.2019.106496}
  {\path{doi:10.1016/j.elecom.2019.106496}}.

\bibitem{Agruss1962a}
B.~Agruss, H.~R. Karas, V.~L. Decker, Design and development of a liquid metal
  fuel cell, Tech. Rep. ASD-TDR-62-1045, {General Motors Corporation} (1962).

\bibitem{Cairns1967}
E.~J. Cairns, C.~E. Crouthamel, A.~K. Fischer, M.~S. Foster, J.~C. Hesson,
  C.~E. Johnson, H.~Shimotake, A.~D. Tevebaugh, Galvanic {{Cells}} with
  {{Fused}}-{{Salt Electrolytes}}, {{ANL}}-7316, {Argonne National Laboratory},
  1967.

\bibitem{Vogel1967}
R.~C. Vogel, M.~Levenson, E.~R. Proud, J.~Royal, Chemical engineering division
  research highlights, Tech. Rep. ANL-7350, {Argonne National Laboratory}
  (1967).

\bibitem{Foster1967b}
M.~S. Foster, Laboratory {{Studies}} of {{Intermetallic Cells}}, in: C.~E.
  Crouthamel, H.~L. Recht (Eds.), Regenerative {{EMF Cells}}, {American
  Chemical Society}, 1967, pp. 136--148.

\bibitem{Ashour2017a}
R.~Ashour, D.~H. Kelley, A.~Salas, M.~Starace, N.~Weber, T.~Weier, Competing
  forces in liquid metal electrodes and batteries, J. Power Sources 378 (2018)
  301--310.
\newblock \href {https://doi.org/10.1016/j.jpowsour.2017.12.042}
  {\path{doi:10.1016/j.jpowsour.2017.12.042}}.

\bibitem{Weber2018}
N.~Weber, M.~Nimtz, P.~Personnettaz, A.~Salas, T.~Weier, Electromagnetically
  driven convection suitable for mass transfer enhancement in liquid metal
  batteries, Appl. Therm. Eng. 143 (2018) 293--301.
\newblock \href {https://doi.org/10.1016/j.applthermaleng.2018.07.067}
  {\path{doi:10.1016/j.applthermaleng.2018.07.067}}.

\bibitem{Herreman2020a}
W.~Herreman, S.~B{\'e}nard, C.~Nore, P.~Personnettaz, L.~Cappanera, J.-L.
  Guermond, Solutal buoyancy and electrovortex flow in liquid metal batteries,
  Phys. Rev. Fluids 5~(7) (2020) 074501.
\newblock \href {https://doi.org/10.1103/PhysRevFluids.5.074501}
  {\path{doi:10.1103/PhysRevFluids.5.074501}}.

\bibitem{Weber2020}
N.~Weber, M.~Nimtz, P.~Personnettaz, T.~Weier, D.~Sadoway, Numerical simulation
  of mass transfer enhancement in liquid metal batteries by means of
  electro-vortex flow, J. Power Sources Adv. 1 (2020) 100004.
\newblock \href {https://doi.org/10.1016/j.powera.2020.100004}
  {\path{doi:10.1016/j.powera.2020.100004}}.

\bibitem{Herreman2021}
W.~Herreman, C.~Nore, L.~Cappanera, J.-L. Guermond, Efficient mixing by
  swirling electrovortex flows in liquid metal batteries, J. Fluid Mech. 915
  (2021) A17.
\newblock \href {https://doi.org/10.1017/jfm.2021.79}
  {\path{doi:10.1017/jfm.2021.79}}.

\bibitem{Kelley2014}
D.~H. Kelley, D.~R. Sadoway, Mixing in a liquid metal electrode, Phys. Fluids
  26~(5) (2014) 057102.
\newblock \href {https://doi.org/10.1063/1.4875815}
  {\path{doi:10.1063/1.4875815}}.

\bibitem{Beltran2016}
A.~Beltr{\'a}n, {{MHD}} natural convection flow in a liquid metal electrode,
  Appl. Therm. Eng. 114 (2016) 1203--1212.
\newblock \href {https://doi.org/10.1016/j.applthermaleng.2016.09.006}
  {\path{doi:10.1016/j.applthermaleng.2016.09.006}}.

\bibitem{Shen2015}
Y.~Shen, O.~Zikanov, Thermal convection in a liquid metal battery, Theor.
  Comput. Fluid Dyn. 30~(4) (2016) 275--294.
\newblock \href {https://doi.org/10.1007/s00162-015-0378-1}
  {\path{doi:10.1007/s00162-015-0378-1}}.

\bibitem{Koellner2017}
T.~K{\"o}llner, T.~Boeck, J.~Schumacher, Thermal {{Rayleigh}}-{{Marangoni}}
  convection in a three-layer liquid-metal-battery model, Phys. Rev. E 95
  (2017) 053114.
\newblock \href {https://doi.org/10.1103/PhysRevE.95.053114}
  {\path{doi:10.1103/PhysRevE.95.053114}}.

\bibitem{Personnettaz2018a}
P.~Personnettaz, P.~Beckstein, S.~Landgraf, T.~K{\"o}llner, M.~Nimtz, N.~Weber,
  T.~Weier, Thermally driven convection in {{Li}}\$||\${{Bi}} liquid metal
  batteries, J. Power Sources 401 (2018) 362--374.
\newblock \href {https://doi.org/10.1016/j.jpowsour.2018.08.069}
  {\path{doi:10.1016/j.jpowsour.2018.08.069}}.

\bibitem{Stefani2015}
F.~Stefani, V.~Galindo, C.~Kasprzyk, S.~Landgraf, M.~Seilmayer, M.~Starace,
  N.~Weber, T.~Weier, Magnetohydrodynamic effects in liquid metal batteries,
  IOP Conf. Ser. Mater. Sci. Eng. 143 (2016) 012024.
\newblock \href {https://doi.org/10.1088/1757-899X/143/1/012024}
  {\path{doi:10.1088/1757-899X/143/1/012024}}.

\bibitem{Kim2013b}
H.~Kim, D.~A. Boysen, J.~M. Newhouse, B.~L. Spatocco, B.~Chung, P.~J. Burke,
  D.~J. Bradwell, K.~Jiang, A.~A. Tomaszowska, K.~Wang, W.~Wei, L.~A. Ortiz,
  S.~A. Barriga, S.~M. Poizeau, D.~R. Sadoway, Liquid {{Metal Batteries}}:
  {{Past}}, {{Present}}, and {{Future}}, Chem. Rev. 113~(3) (2013) 2075--2099.
\newblock \href {https://doi.org/10.1021/cr300205k}
  {\path{doi:10.1021/cr300205k}}.

\bibitem{Lawroski1962b}
S.~Lawroski, R.~C. Vogel, V.~H. Munnecke, Chemical engineering division summary
  report, Tech. Rep. ANL-6543, {Argonne National Laboratory} (1962).

\bibitem{Crouthamel1967}
C.~E. Crouthamel, H.~L. Recht (Eds.), Regenerative {{EMF Cells}}, Vol.~64,
  {American Chemical Society}, 1967.

\bibitem{Shimotake1967}
H.~Shimotake, E.~J. Cairns, Bimetallic galvanic cells with fused-salt
  electrolytes, in: Advances in {{Energy Conversion Engineering}}, {ASME},
  {Florida}, 1967, pp. 951--962.

\bibitem{Bradwell2011}
D.~J. Bradwell, Liquid {{Metal Batteries}}: {{Ambipolar Electrolysis}} and
  {{Alkaline Earth Electroalloying Cells}}, Ph.D. thesis, Massachusetts
  Institute of Technology (2011).

\bibitem{Ouchi2014}
T.~Ouchi, H.~Kim, X.~Ning, D.~R. Sadoway, Calcium-{{Antimony Alloys}} as
  {{Electrodes}} for {{Liquid Metal Batteries}}, J. Electrochem. Soc. 161~(12)
  (2014) A1898--A1904.

\bibitem{Ouchi2016}
T.~Ouchi, H.~Kim, B.~L. Spatocco, D.~R. Sadoway, Calcium-based multi-element
  chemistry for grid-scale electrochemical energy storage, Nat. Commun. 7
  (2016) 10999.
\newblock \href {https://doi.org/10.1038/ncomms10999}
  {\path{doi:10.1038/ncomms10999}}.

\bibitem{Blanchard2013}
A.~Blanchard, Enabling multi-cation electrolyte usage in {{LMBs}} for lower
  cost and operating temperature, Ph.D. thesis, Massachusetts Institute of
  Technology (2013).

\bibitem{Agruss1963}
B.~Agruss, The {{Thermally Regenerative Liquid}}-{{Metal Cell}}, J.
  Electrochem. Soc. 110~(11) (1963) 1097--1103.

\bibitem{bird2002transportphenomena}
R.~B. Bird, W.~E. Stewart, E.~N. Lightfoot, Transport Phenomena, Wiley, 2002.
\newblock \href {https://doi.org/10.1115/1.1424298}
  {\path{doi:10.1115/1.1424298}}.

\bibitem{weber2021cell}
N.~Weber, C.~Duczek, G.~M. Horstmann, S.~Landgraf, M.~Nimtz, P.~Personnettaz,
  T.~Weier, D.~R. Sadoway, Cell voltage model for {Li}-{Bi} liquid metal
  batteries, Appl. Energy, under revision (2021).

\bibitem{leal2007advanced}
L.~G. Leal, Advanced Transport Phenomena: Fluid Mechanics and Convective
  Transport Processes, Cambridge Series in Chemical Engineering, Cambridge
  University Press, 2007.
\newblock \href {https://doi.org/10.1017/CBO9780511800245}
  {\path{doi:10.1017/CBO9780511800245}}.

\bibitem{Fazio2015}
C.~Fazio, Handbook on {{Lead}}-bismuth {{Eutectic Alloy}} and {{Lead
  Properties}}, {{Materials Compatibility}}, {{Thermal}}-hydraulics and
  {{Technologies}}, Tech. Rep. 7268, {Nuclear Energy Agency} (2015).

\bibitem{Iida2015}
T.~Iida, R.~I.~L. Guthrie, The Thermophysical Properties of Metallic Liquids,
  {Oxford University Press}, {United Kingdom}, 2015.

\bibitem{Roy1988}
A.~K. Roy, R.~P. Chhabra, Prediction of {{Solute Diffusion Coefficients}} in
  {{Liquid Metals}}, Metall. Trans. A 19 (1988) 273--279.

\bibitem{Sparrow1964}
E.~M. Sparrow, R.~J. Goldstein, V.~K. Jonsson, Thermal instability in a
  horizontal fluid layer: Effect of boundary conditions and non-linear
  temperature profile, J. Fluid Mech. 18~(4) (1964) 513--528.
\newblock \href {https://doi.org/10.1017/S0022112064000386}
  {\path{doi:10.1017/S0022112064000386}}.

\bibitem{KulackiGoldstein:1972}
F.~A. Kulacki, R.~J. Goldstein, Thermal convection in a horizontal fluid layer
  with uniform volumetric energy sources, J. Fluid Mech. 55~(02) (1972)
  271--287.

\bibitem{kulacki1975hydrodynamic}
F.~A. Kulacki, R.~J. Goldstein, Hydrodynamic instability in fluid layers with
  uniform volumetric energy sources, Appl. Sci. Res. 31~(2) (1975) 81--109.
\newblock \href {https://doi.org/10.1007/bf01795829}
  {\path{doi:10.1007/bf01795829}}.

\bibitem{jasak2020practical}
H.~Jasak, T.~Uroi{\'c}, Practical Computational Fluid Dynamics with the Finite
  Volume Method, Springer International Publishing, 2020, pp. 103--161.
\newblock \href {https://doi.org/10.1007/978-3-030-37518-8_4}
  {\path{doi:10.1007/978-3-030-37518-8_4}}.

\bibitem{Weller1998}
H.~G. Weller, G.~Tabor, H.~Jasak, C.~Fureby, A tensorial approach to
  computational continuum mechanics using object-oriented techniques, Comput.
  Phys. 12~(6) (1998) 620--631.

\bibitem{Beale2016}
S.~B. Beale, H.-W. Choi, J.~G. Pharoah, H.~K. Roth, H.~Jasak, D.~H. Jeon,
  Open-source computational model of a solid oxide fuel cell, Comput. Phys.
  Commun. 200 (2016) 15--26.
\newblock \href {https://doi.org/10.1016/j.cpc.2015.10.007}
  {\path{doi:10.1016/j.cpc.2015.10.007}}.

\bibitem{Weber2017b}
N.~Weber, P.~Beckstein, V.~Galindo, M.~Starace, T.~Weier, Electro-vortex flow
  simulation using coupled meshes, Comput. Fluids 168 (2018) 101--109.
\newblock \href {https://doi.org/10.1016/j.compfluid.2018.03.047}
  {\path{doi:10.1016/j.compfluid.2018.03.047}}.

\bibitem{tukovic2012moving}
\v{Z}. Tukovi{\'c}, H.~Jasak, A moving mesh finite volume interface tracking
  method for surface tension dominated interfacial fluid flow, Comput. Fluids
  55 (2012) 70--84.
\newblock \href {https://doi.org/10.1016/j.compfluid.2011.11.003}
  {\path{doi:10.1016/j.compfluid.2011.11.003}}.

\bibitem{Foster1964}
M.~S. Foster, S.~E. Wood, C.~E. Crouthamel, Thermodynamics of {{Binary
  Alloys}}. {{I}}. {{The Lithium}}-{{Bismuth System}}, Inorg. Chem. 3~(10)
  (1964) 1428--1431.

\bibitem{Ohse1985}
R.~W. Ohse (Ed.), Handbook of Thermodynamic and Transport Properties of Alkali
  Metals, no.~30 in {{IUPAC}} Chemical Data {{Series}}, {Blackwell Scientific},
  {Oxford}, 1985.

\bibitem{Raseman1960}
C.~J. Raseman, H.~Susskind, G.~Farber, W.~McNulty, F.~Salzano, Engineering
  experience at {{Brookhaven National Laboratory}} in handling fused chloride
  salts, Tech. Rep. BNL 627, {Brookhaven National Laboratory} (1960).

\bibitem{Janz1979}
G.~J. Janz, R.~P.~T. Tomkins, C.~B. Allen, Molten {{Salts}}: {{Volume}} 4,
  {{Part}} 4 {{Mixed Halide Melts Electrical Conductance}}, {{Density}},
  {{Viscosity}}, and {{Surface Tension Data}}, J. Phys. Chem. Ref. Data 8~(1)
  (1979) 125--302.
\newblock \href {https://doi.org/10.1063/1.555590}
  {\path{doi:10.1063/1.555590}}.

\bibitem{Janz1988}
G.~J. Janz, Thermodynamic and transport properties for molton salts:
  {{Correlation}} equations for critically evaluated density, surface tension,
  electrical conductance, and viscosity data, J. Phys. Chem. Ref. Data 17
  (1988).

\bibitem{linstrom2018nist}
P.~J. Linstrom, W.~G. Mallard, NIST Chemistry WebBook, NIST Standard Reference
  Database Number 69, National Institute of Standards and Technology,
  Gaithersburg MD, 20899, 2018.
\newblock \href {https://doi.org/10.18434/T4D30} {\path{doi:10.18434/T4D30}}.

\bibitem{Koellner2014}
T.~K{\"o}llner, M.~Rossi, F.~Broer, T.~Boeck, Chemical convection in the
  methylene-blue\textendash glucose system: {{Optimal}} perturbations and
  three-dimensional simulations, Phys. Rev. E 90~(5) (2014) 053004.
\newblock \href {https://doi.org/10.1103/PhysRevE.90.053004}
  {\path{doi:10.1103/PhysRevE.90.053004}}.

\bibitem{Blackburn2019}
H.~Blackburn, D.~Lee, T.~Albrecht, J.~Singh, Semtex: {{A}} spectral
  element\textendash{{Fourier}} solver for the incompressible
  {{Navier}}\textendash{{Stokes}} equations in cylindrical or {{Cartesian}}
  coordinates, Comput. Phys. Commun. 245 (2019) 106804.
\newblock \href {https://doi.org/10.1016/j.cpc.2019.05.015}
  {\path{doi:10.1016/j.cpc.2019.05.015}}.

\bibitem{goldstein1995onset}
R.~J. Goldstein, R.~J. Volino, Onset and development of natural convection
  above a suddenly heated horizontal surface, J. Heat Transf. 117~(4) (1995)
  808--821.
\newblock \href {https://doi.org/10.1115/1.2836296}
  {\path{doi:10.1115/1.2836296}}.

\bibitem{Philippi2019}
J.~Philippi, M.~Berhanu, J.~Derr, S.~{Courrech du Pont}, Solutal convection
  induced by dissolution, Phys. Rev. Fluids 4~(10) (2019) 103801.
\newblock \href {https://doi.org/10.1103/PhysRevFluids.4.103801}
  {\path{doi:10.1103/PhysRevFluids.4.103801}}.

\bibitem{Backhaus2011}
S.~Backhaus, K.~Turitsyn, R.~E. Ecke, Convective {{Instability}} and {{Mass
  Transport}} of {{Diffusion Layers}} in a {{Hele}}-{{Shaw Geometry}}, Phys.
  Rev. Lett. 106~(10) (2011) 104501.
\newblock \href {https://doi.org/10.1103/PhysRevLett.106.104501}
  {\path{doi:10.1103/PhysRevLett.106.104501}}.

\bibitem{prakash1997flow}
A.~Prakash, K.~Yasuda, F.~Otsubo, K.~Kuwahara, T.~Doi, Flow coupling mechanisms
  in two-layer {Rayleigh}-{Benard} convection, Experiments in Fluids 23~(3)
  (1997) 252--261.
\newblock \href {https://doi.org/10.1007/s003480050108}
  {\path{doi:10.1007/s003480050108}}.

\bibitem{2016}
Springer Handbook of Electrochemical Energy, 1st Edition, {Springer Berlin
  Heidelberg}, {New York, NY}, 2016.

\bibitem{Goluskin2016}
D.~Goluskin, Internally {{Heated Convection}} and {{Rayleigh}}-{{B\'enard
  Convection}}, {Springer International Publishing}, {Cham}, 2016.
\newblock \href {https://doi.org/10.1007/978-3-319-23941-5}
  {\path{doi:10.1007/978-3-319-23941-5}}.

\bibitem{JohnsonNarayanan:1997}
D.~Johnson, R.~Narayanan, Geometric effects on convective coupling and
  interfacial structures in bilayer convection, Phys. Rev. E 56~(5) (1997)
  5462--5472.

\bibitem{PeckoverHutchinson:1974}
R.~S. Peckover, I.~H. Hutchinson, Convective rolls driven by internal heat
  sources, Phys. Fluids 17~(7) (1974) 1369--1371.

\bibitem{Veronis:1963}
G.~Veronis, Penetrative convection, Astrophys. J. 137 (1963) 641--663.

\bibitem{JahnReineke:1974}
M.~Jahn, H.~H. Reineke, Free convection heat transfer with internal heat
  sources calculations and measurements, in: Proc. {International Heat Transfer
  Conference} 5, Tokyo, 1974, pp. 74--78.

\bibitem{RalphMcGreevyPeckover:1976}
J.~C. Ralph, R.~McGreevy, R.~S. Peckover, Experiments in turbulent thermal
  convection driven by internal heat sources, in: Heat Transfer and Turbulent
  Buoyant Convection, Vol.~2, McGraw-Hill, 1976, pp. 587--599.

\bibitem{Tveitereid:1978}
M.~Tveitereid, Thermal convection in a horizontal fluid layer with internal
  heat sources, Int. J. Heat Mass Transf. 21 (1978) 335--339.

\bibitem{KulackiRichards:1985}
F.~A. Kulacki, D.~E. Richards, Natural convection in plane layers and cavities
  with volumetric energy sources, in: S.~Kakac, W.~Aung, R.~Viskanta (Eds.),
  Natural Convection: Fundamentals and Applications, Hemisphere, Washington,
  1985, pp. 179--255.

\end{thebibliography}
\end{document}